\RequirePackage{lineno}
\documentclass[traditabstract,referee,longauth]{aa}
\usepackage{longtable,lscape}
\usepackage{amsmath}
\usepackage{amssymb}
\usepackage{newcent}
\usepackage{url,hyperref}
\usepackage{txfonts}
\usepackage{graphicx}
\usepackage{natbib}
\usepackage{subfigure}
\usepackage{footmisc}

\makeatletter
\renewcommand*{\@fnsymbol}[1]{\ifcase#1\or*\or$\dagger$\or$\ddagger$\or**\or$\dagger\dagger$\or$\ddagger\ddagger$\fi}
\makeatother

\usepackage{color}

\bibpunct{(}{)}{;}{a}{}{,} 

\newcommand{\hess}{\textsc{H.E.S.S.}}

\newcommand{\swift}{\textsl{Swift}}
\newcommand{\swiftxrt}{\textsl{Swift}-XRT}
\newcommand{\swiftuvot}{\textsl{Swift}-UVOT}
\newcommand{\fer}{{\sl {\it Fermi}}}
\newcommand{\fla}{\fer-LAT}
\newcommand{\nustar}{\textsl{NuSTAR}}

\newcommand{\dgr}{\ensuremath{^\circ}}

\newcommand{\ergs}{erg\,s$^{-1}$}%

\newcommand{\gr}{$\gamma$-ray}
\newcommand{\grs}{$\gamma$-rays}

\newcommand{\pks}{PKS~2155-304}

\begin{document}
\title{Simultaneous observations of the blazar \pks\ from Ultra-Violet to TeV energies.}

\author{
H.~Abdalla \inst{\ref{NWU}}
\and R.~Adam \inst{\ref{LLR}}
\and F.~Aharonian \inst{\ref{MPIK},\ref{DIAS},\ref{RAU}}
\and F.~Ait~Benkhali \inst{\ref{MPIK}}
\and E.O.~Ang\"uner \inst{\ref{CPPM}}
\and M.~Arakawa \inst{\ref{Rikkyo}}
\and C.~Arcaro \inst{\ref{NWU}}
\and C.~Armand \inst{\ref{LAPP}}
\and H.~Ashkar \inst{\ref{IRFU}}
\and M.~Backes \inst{\ref{UNAM},\ref{NWU}}
\and V.~Barbosa~Martins \inst{\ref{DESY}}
\and M.~Barnard \inst{\ref{NWU}}
\and Y.~Becherini \inst{\ref{Linnaeus}}
\and D.~Berge \inst{\ref{DESY}}
\and K.~Bernl\"ohr \inst{\ref{MPIK}}
\and R.~Blackwell \inst{\ref{Adelaide}}
\and M.~B\"ottcher \inst{\ref{NWU}}
\and C.~Boisson \inst{\ref{LUTH}}
\and J.~Bolmont \inst{\ref{LPNHE}}
\and S.~Bonnefoy \inst{\ref{DESY}}
\and J.~Bregeon \inst{\ref{LUPM}}
\and M.~Breuhaus \inst{\ref{MPIK}}
\and F.~Brun \inst{\ref{IRFU}}
\and P.~Brun \inst{\ref{IRFU}}
\and M.~Bryan \inst{\ref{GRAPPA}}
\and M.~B\"{u}chele \inst{\ref{ECAP}}
\and T.~Bulik \inst{\ref{UWarsaw}}
\and T.~Bylund \inst{\ref{Linnaeus}}
\and S.~Caroff \inst{\ref{LPNHE}}
\and A.~Carosi \inst{\ref{LAPP}}
\and S.~Casanova \inst{\ref{IFJPAN},\ref{MPIK}}
\and M.~Cerruti \inst{\ref{LPNHE},\ref{CerrutiNowAt}}  \protect\footnotemark[1] 
\and T.~Chand \inst{\ref{NWU}}
\and S.~Chandra \inst{\ref{NWU}}
\and A.~Chen \inst{\ref{WITS}}
\and S.~Colafrancesco \inst{\ref{WITS}} 
\and M.~Cury{\l}o \inst{\ref{UWarsaw}}
\and I.D.~Davids \inst{\ref{UNAM}}
\and C.~Deil \inst{\ref{MPIK}}
\and J.~Devin \inst{\ref{CENBG}}
\and P.~deWilt \inst{\ref{Adelaide}}
\and L.~Dirson \inst{\ref{HH}}
\and A.~Djannati-Ata\"i \inst{\ref{APC}}
\and A.~Dmytriiev \inst{\ref{LUTH}}
\and A.~Donath \inst{\ref{MPIK}}
\and V.~Doroshenko \inst{\ref{IAAT}}
\and J.~Dyks \inst{\ref{NCAC}}
\and K.~Egberts \inst{\ref{UP}}
\and G.~Emery \inst{\ref{LPNHE}}
\and J.-P.~Ernenwein \inst{\ref{CPPM}}
\and S.~Eschbach \inst{\ref{ECAP}}
\and K.~Feijen \inst{\ref{Adelaide}}
\and S.~Fegan \inst{\ref{LLR}}
\and A.~Fiasson \inst{\ref{LAPP}}
\and G.~Fontaine \inst{\ref{LLR}}
\and S.~Funk \inst{\ref{ECAP}}
\and M.~F\"u{\ss}ling \inst{\ref{DESY}}
\and S.~Gabici \inst{\ref{APC}}
\and Y.A.~Gallant \inst{\ref{LUPM}}
\and F.~Gat{\'e} \inst{\ref{LAPP}}
\and G.~Giavitto \inst{\ref{DESY}}
\and L.~Giunti \inst{\ref{APC}}
\and D.~Glawion \inst{\ref{LSW}}
\and J.F.~Glicenstein \inst{\ref{IRFU}}
\and D.~Gottschall \inst{\ref{IAAT}}
\and M.-H.~Grondin \inst{\ref{CENBG}}
\and J.~Hahn \inst{\ref{MPIK}}
\and M.~Haupt \inst{\ref{DESY}}
\and G.~Heinzelmann \inst{\ref{HH}}
\and G.~Henri \inst{\ref{Grenoble}}
\and G.~Hermann \inst{\ref{MPIK}}
\and J.A.~Hinton \inst{\ref{MPIK}}
\and W.~Hofmann \inst{\ref{MPIK}}
\and C.~Hoischen \inst{\ref{UP}}
\and T.~L.~Holch \inst{\ref{HUB}}
\and M.~Holler \inst{\ref{LFUI}}
\and D.~Horns \inst{\ref{HH}}
\and D.~Huber \inst{\ref{LFUI}}
\and H.~Iwasaki \inst{\ref{Rikkyo}}
\and M.~Jamrozy \inst{\ref{UJK}}
\and D.~Jankowsky \inst{\ref{ECAP}}
\and F.~Jankowsky \inst{\ref{LSW}}
\and A.~Jardin-Blicq \inst{\ref{MPIK}}
\and I.~Jung-Richardt \inst{\ref{ECAP}}
\and M.A.~Kastendieck \inst{\ref{HH}}
\and K.~Katarzy{\'n}ski \inst{\ref{NCUT}}
\and M.~Katsuragawa \inst{\ref{KAVLI}}
\and U.~Katz \inst{\ref{ECAP}}
\and D.~Khangulyan \inst{\ref{Rikkyo}}
\and B.~Kh\'elifi \inst{\ref{APC}}
\and J.~King \inst{\ref{LSW}}
\and S.~Klepser \inst{\ref{DESY}}
\and W.~Klu\'{z}niak \inst{\ref{NCAC}}
\and Nu.~Komin \inst{\ref{WITS}}
\and K.~Kosack \inst{\ref{IRFU}}
\and D.~Kostunin \inst{\ref{DESY}} 
\and M.~Kreter \inst{\ref{NWU}}
\and G.~Lamanna \inst{\ref{LAPP}}
\and A.~Lemi\`ere \inst{\ref{APC}}
\and M.~Lemoine-Goumard \inst{\ref{CENBG}}
\and J.-P.~Lenain \inst{\ref{LPNHE}}
\and E.~Leser \inst{\ref{UP},\ref{DESY}}
\and C.~Levy \inst{\ref{LPNHE}}
\and T.~Lohse \inst{\ref{HUB}}
\and I.~Lypova \inst{\ref{DESY}}
\and J.~Mackey \inst{\ref{DIAS}}
\and J.~Majumdar \inst{\ref{DESY}}
\and D.~Malyshev \inst{\ref{IAAT}}
\and V.~Marandon \inst{\ref{MPIK}}
\and A.~Marcowith \inst{\ref{LUPM}}
\and A.~Mares \inst{\ref{CENBG}}
\and C.~Mariaud \inst{\ref{LLR}}
\and G.~Mart\'i-Devesa \inst{\ref{LFUI}}
\and R.~Marx \inst{\ref{MPIK}}
\and G.~Maurin \inst{\ref{LAPP}}
\and P.J.~Meintjes \inst{\ref{UFS}}
\and A.M.W.~Mitchell \inst{\ref{MPIK},\ref{MitchellNowAt}}
\and R.~Moderski \inst{\ref{NCAC}}
\and M.~Mohamed \inst{\ref{LSW}}
\and L.~Mohrmann \inst{\ref{ECAP}}
\and C.~Moore \inst{\ref{Leicester}}
\and E.~Moulin \inst{\ref{IRFU}}
\and J.~Muller \inst{\ref{LLR}}
\and T.~Murach \inst{\ref{DESY}}
\and S.~Nakashima  \inst{\ref{RIKKEN}}
\and M.~de~Naurois \inst{\ref{LLR}}
\and H.~Ndiyavala  \inst{\ref{NWU}}
\and F.~Niederwanger \inst{\ref{LFUI}}
\and J.~Niemiec \inst{\ref{IFJPAN}}
\and L.~Oakes \inst{\ref{HUB}}
\and P.~O'Brien \inst{\ref{Leicester}}
\and H.~Odaka \inst{\ref{Tokyo}}
\and S.~Ohm \inst{\ref{DESY}}
\and E.~de~Ona~Wilhelmi \inst{\ref{DESY}}
\and M.~Ostrowski \inst{\ref{UJK}}
\and I.~Oya \inst{\ref{DESY}}
\and M.~Panter \inst{\ref{MPIK}}
\and R.D.~Parsons \inst{\ref{MPIK}}
\and C.~Perennes \inst{\ref{LPNHE}}
\and P.-O.~Petrucci \inst{\ref{Grenoble}}
\and B.~Peyaud \inst{\ref{IRFU}}
\and Q.~Piel \inst{\ref{LAPP}}
\and S.~Pita \inst{\ref{APC}}
\and V.~Poireau \inst{\ref{LAPP}}
\and A.~Priyana~Noel \inst{\ref{UJK}}
\and D.A.~Prokhorov \inst{\ref{WITS}}
\and H.~Prokoph \inst{\ref{DESY}}
\and G.~P\"uhlhofer \inst{\ref{IAAT}}
\and M.~Punch \inst{\ref{APC},\ref{Linnaeus}}
\and A.~Quirrenbach \inst{\ref{LSW}}
\and S.~Raab \inst{\ref{ECAP}}
\and R.~Rauth \inst{\ref{LFUI}}
\and A.~Reimer \inst{\ref{LFUI}}
\and O.~Reimer \inst{\ref{LFUI}}
\and Q.~Remy \inst{\ref{LUPM}}
\and M.~Renaud \inst{\ref{LUPM}}
\and F.~Rieger \inst{\ref{MPIK}}
\and L.~Rinchiuso \inst{\ref{IRFU}}
\and C.~Romoli \inst{\ref{MPIK}}  \protect\footnotemark[1] 
\and G.~Rowell \inst{\ref{Adelaide}}
\and B.~Rudak \inst{\ref{NCAC}}
\and E.~Ruiz-Velasco \inst{\ref{MPIK}}
\and V.~Sahakian \inst{\ref{YPI}}
\and S.~Sailer \inst{\ref{MPIK}}
\and S.~Saito \inst{\ref{Rikkyo}}
\and D.A.~Sanchez \inst{\ref{LAPP}} \protect\footnotemark[1] 
\and A.~Santangelo \inst{\ref{IAAT}}
\and M.~Sasaki \inst{\ref{ECAP}}
\and R.~Schlickeiser \inst{\ref{RUB}}
\and F.~Sch\"ussler \inst{\ref{IRFU}}
\and A.~Schulz \inst{\ref{DESY}}
\and H.M.~Schutte \inst{\ref{NWU}}
\and U.~Schwanke \inst{\ref{HUB}}
\and S.~Schwemmer \inst{\ref{LSW}}
\and M.~Seglar-Arroyo \inst{\ref{IRFU}}
\and M.~Senniappan \inst{\ref{Linnaeus}}
\and A.S.~Seyffert \inst{\ref{NWU}}
\and N.~Shafi \inst{\ref{WITS}}
\and K.~Shiningayamwe \inst{\ref{UNAM}}
\and R.~Simoni \inst{\ref{GRAPPA}}
\and A.~Sinha \inst{\ref{APC}}
\and H.~Sol \inst{\ref{LUTH}}
\and A.~Specovius \inst{\ref{ECAP}}
\and M.~Spir-Jacob \inst{\ref{APC}}
\and {\L.}~Stawarz \inst{\ref{UJK}}
\and R.~Steenkamp \inst{\ref{UNAM}}
\and C.~Stegmann \inst{\ref{UP},\ref{DESY}}
\and C.~Steppa \inst{\ref{UP}}
\and T.~Takahashi  \inst{\ref{KAVLI}}
\and T.~Tavernier \inst{\ref{IRFU}}
\and A.M.~Taylor \inst{\ref{DESY}}
\and R.~Terrier \inst{\ref{APC}}
\and D.~Tiziani \inst{\ref{ECAP}}
\and M.~Tluczykont \inst{\ref{HH}}
\and C.~Trichard \inst{\ref{LLR}}
\and M.~Tsirou \inst{\ref{LUPM}}
\and N.~Tsuji \inst{\ref{Rikkyo}}
\and R.~Tuffs \inst{\ref{MPIK}}
\and Y.~Uchiyama \inst{\ref{Rikkyo}}
\and D.J.~van~der~Walt \inst{\ref{NWU}}
\and C.~van~Eldik \inst{\ref{ECAP}}
\and C.~van~Rensburg \inst{\ref{NWU}}
\and B.~van~Soelen \inst{\ref{UFS}}
\and G.~Vasileiadis \inst{\ref{LUPM}}
\and J.~Veh \inst{\ref{ECAP}}
\and C.~Venter \inst{\ref{NWU}}
\and P.~Vincent \inst{\ref{LPNHE}}
\and J.~Vink \inst{\ref{GRAPPA}}
\and H.J.~V\"olk \inst{\ref{MPIK}}
\and T.~Vuillaume \inst{\ref{LAPP}} \protect\footnotemark[1] 
\and Z.~Wadiasingh \inst{\ref{NWU}}
\and S.J.~Wagner \inst{\ref{LSW}}
\and R.~White \inst{\ref{MPIK}}
\and A.~Wierzcholska \inst{\ref{IFJPAN},\ref{LSW}}
\and R.~Yang \inst{\ref{MPIK}}
\and H.~Yoneda \inst{\ref{KAVLI}}
\and M.~Zacharias \inst{\ref{NWU}}
\and R.~Zanin \inst{\ref{MPIK}}
\and A.A.~Zdziarski \inst{\ref{NCAC}}
\and A.~Zech \inst{\ref{LUTH}}
\and J.~Zorn \inst{\ref{MPIK}}
\and N.~\.Zywucka \inst{\ref{NWU}}
\\ And\\
 G.~M.~Madejski\inst{\ref{KIPAC}} \protect\footnotemark[1] 
\and K.~Nalewajko\inst{\ref{NCAC}}
\and K.~K.~Madsen \inst{\ref{Cahill}}
\and J.~Chiang\inst{\ref{KIPAC}}
\and M.~Balokovi\'{c} \inst{\ref{HSCA},\ref{BHIHU}}
\and D.~Paneque \inst{\ref{MPIM}}
\and A.~K.~Furniss \inst{\ref{CSU}}
\and M.~Hayashida \inst{\ref{Rikkyo}}
\and C.~M.~Urry\inst{\ref{Yale}}
\and M.~Ajello \inst{\ref{DPA}}
\and F.~A.~Harrison \inst{\ref{Cahill}}
\and B. Giebels\inst{\ref{LLR}}
\and D.~Stern \inst{\ref{JPL}}
\and K.~Forster \inst{\ref{Cahill}}
\and P.~Giommi\inst{\ref{ASI}}
\and M.~Perri\inst{\ref{ASI},\ref{INAF}} \protect\footnotemark[1] 
\and S.~Puccetti\inst{\ref{ASI}}
\and A. ~Zoglauer \inst{\ref{SSL}}
\and G.~Tagliaferri  \inst{\ref{INAFB}}
}

\institute{
Centre for Space Research, North-West University, Potchefstroom 2520, South Africa \label{NWU} \and 
Universit\"at Hamburg, Institut f\"ur Experimentalphysik, Luruper Chaussee 149, D 22761 Hamburg, Germany \label{HH} \and 
Max-Planck-Institut f\"ur Kernphysik, P.O. Box 103980, D 69029 Heidelberg, Germany \label{MPIK} \and 
Dublin Institute for Advanced Studies, 31 Fitzwilliam Place, Dublin 2, Ireland \label{DIAS} \and 
High Energy Astrophysics Laboratory, RAU,  123 Hovsep Emin St  Yerevan 0051, Armenia \label{RAU} \and
Yerevan Physics Institute, 2 Alikhanian Brothers St., 375036 Yerevan, Armenia \label{YPI} \and
Institut f\"ur Physik, Humboldt-Universit\"at zu Berlin, Newtonstr. 15, D 12489 Berlin, Germany \label{HUB} \and
University of Namibia, Department of Physics, Private Bag 13301, Windhoek, Namibia, 12010 \label{UNAM} \and
GRAPPA, Anton Pannekoek Institute for Astronomy, University of Amsterdam,  Science Park 904, 1098 XH Amsterdam, The Netherlands \label{GRAPPA} \and
Department of Physics and Electrical Engineering, Linnaeus University,  351 95 V\"axj\"o, Sweden \label{Linnaeus} \and
Institut f\"ur Theoretische Physik, Lehrstuhl IV: Weltraum und Astrophysik, Ruhr-Universit\"at Bochum, D 44780 Bochum, Germany \label{RUB} \and
Institut f\"ur Astro- und Teilchenphysik, Leopold-Franzens-Universit\"at Innsbruck, A-6020 Innsbruck, Austria \label{LFUI} \and
School of Physical Sciences, University of Adelaide, Adelaide 5005, Australia \label{Adelaide} \and
LUTH, Observatoire de Paris, PSL Research University, CNRS, Universit\'e Paris Diderot, 5 Place Jules Janssen, 92190 Meudon, France \label{LUTH} \and
Sorbonne Universit\'e, Universit\'e Paris Diderot, Sorbonne Paris Cit\'e, CNRS/IN2P3, Laboratoire de Physique Nucl\'eaire et de Hautes Energies, LPNHE, 4 Place Jussieu, F-75252 Paris, France \label{LPNHE} \and
Laboratoire Univers et Particules de Montpellier, Universit\'e Montpellier, CNRS/IN2P3,  CC 72, Place Eug\`ene Bataillon, F-34095 Montpellier Cedex 5, France \label{LUPM} \and
IRFU, CEA, Universit\'e Paris-Saclay, F-91191 Gif-sur-Yvette, France \label{IRFU} \and
Astronomical Observatory, The University of Warsaw, Al. Ujazdowskie 4, 00-478 Warsaw, Poland \label{UWarsaw} \and
Aix Marseille Universit\'e, CNRS/IN2P3, CPPM, Marseille, France \label{CPPM} \and
Instytut Fizyki J\c{a}drowej PAN, ul. Radzikowskiego 152, 31-342 Krak{\'o}w, Poland \label{IFJPAN} \and
School of Physics, University of the Witwatersrand, 1 Jan Smuts Avenue, Braamfontein, Johannesburg, 2050 South Africa \label{WITS} \and
Laboratoire d'Annecy de Physique des Particules, Univ. Grenoble Alpes, Univ. Savoie Mont Blanc, CNRS, LAPP, 74000 Annecy, France \label{LAPP} \and
Landessternwarte, Universit\"at Heidelberg, K\"onigstuhl, D 69117 Heidelberg, Germany \label{LSW} \and
Universit\'e Bordeaux, CNRS/IN2P3, Centre d'\'Etudes Nucl\'eaires de Bordeaux Gradignan, 33175 Gradignan, France \label{CENBG} \and
Institut f\"ur Astronomie und Astrophysik, Universit\"at T\"ubingen, Sand 1, D 72076 T\"ubingen, Germany \label{IAAT} \and
Laboratoire Leprince-Ringuet, École Polytechnique, CNRS, Institut Polytechnique de Paris, F-91128 Palaiseau, France \label{LLR} \and
APC, AstroParticule et Cosmologie, Universit\'{e} Paris Diderot, CNRS/IN2P3, CEA/Irfu, Observatoire de Paris, Sorbonne Paris Cit\'{e}, 10, rue Alice Domon et L\'{e}onie Duquet, 75205 Paris Cedex 13, France \label{APC} \and
Univ. Grenoble Alpes, CNRS, IPAG, F-38000 Grenoble, France \label{Grenoble} \and
Department of Physics and Astronomy, The University of Leicester, University Road, Leicester, LE1 7RH, United Kingdom \label{Leicester} \and
Nicolaus Copernicus Astronomical Center, Polish Academy of Sciences, ul. Bartycka 18, 00-716 Warsaw, Poland \label{NCAC} \and
Institut f\"ur Physik und Astronomie, Universit\"at Potsdam,  Karl-Liebknecht-Strasse 24/25, D 14476 Potsdam, Germany \label{UP} \and
Friedrich-Alexander-Universit\"at Erlangen-N\"urnberg, Erlangen Centre for Astroparticle Physics, Erwin-Rommel-Str. 1, D 91058 Erlangen, Germany \label{ECAP} \and
DESY, D-15738 Zeuthen, Germany \label{DESY} \and
Obserwatorium Astronomiczne, Uniwersytet Jagiello{\'n}ski, ul. Orla 171, 30-244 Krak{\'o}w, Poland \label{UJK} \and
Centre for Astronomy, Faculty of Physics, Astronomy and Informatics, Nicolaus Copernicus University,  Grudziadzka 5, 87-100 Torun, Poland \label{NCUT} \and
Department of Physics, University of the Free State,  PO Box 339, Bloemfontein 9300, South Africa \label{UFS} \and
Department of Physics, Rikkyo University, 3-34-1 Nishi-Ikebukuro, Toshima-ku, Tokyo 171-8501, Japan \label{Rikkyo} \and
Kavli Institute for the Physics and Mathematics of the Universe (WPI), The University of Tokyo Institutes for Advanced Study (UTIAS), The University of Tokyo, 5-1-5 Kashiwa-no-Ha, Kashiwa, Chiba, 277-8583, Japan \label{KAVLI} \and
Department of Physics, The University of Tokyo, 7-3-1 Hongo, Bunkyo-ku, Tokyo 113-0033, Japan \label{Tokyo} \and
RIKEN, 2-1 Hirosawa, Wako, Saitama 351-0198, Japan \label{RIKKEN} \and
Now at Physik Institut, Universit\"at Z\"urich, Winterthurerstrasse 190, CH-8057 Z\"urich, Switzerland \label{MitchellNowAt} \and
Now at Institut de Ci\`{e}ncies del Cosmos (ICC UB), Universitat de Barcelona (IEEC-UB), Mart\'{i} Franqu\`es 1, E08028 Barcelona, Spain \label{CerrutiNowAt} 
\and 
Kavli Institute for Particle Astrophysics and Cosmology, Department of Physics and SLAC National Accelerator Laboratory, Stanford University, Stanford, CA 94305, USA \label{KIPAC} 
\and
Cahill Center for Astronomy and Astrophysics, Caltech, Pasadena, CA 91125, USA \label{Cahill} 
\and
California State University - East Bay, 25800 Carlos Bee Boulevard, Hayward, CA 94542 \label{CSU} 
\and
Yale Center for Astronomy and Astrophysics, Physics Department, Yale University, PO Box 208120, New Haven, CT 06520-8120, USA \label{Yale} 
\and
Department of Physics and Astronomy, Clemson University, Kinard Lab of Physics, Clemson, SC 29634-0978, USA \label{DPA} 
\and
Jet Propulsion Laboratory, California Institute of Technology, Pasadena, CA 91109, USA \label{JPL} 
\and
Space Science Laboratory, University of California, Berkeley, CA 94720, USA \label{SSL} 
\and
Center for Astrophysics $\vert$ Harvard \& Smithsonian, 60 Garden Street, Cambridge, MA 02138, USA \label{HSCA} 
\and
Black Hole Initiative at Harvard University, 20 Garden Street, Cambridge, MA 02138, USA   \label{BHIHU} 
\and
ASI Science Data Center, Via del Politecnico snc I - 00133, Roma, Italy \label{ASI} 
\and
INAF - Osservatorio Astronomico di Roma, via di Frascati 33, I - 00040 Monteporzio, Italy \label{INAF} 
\and
INAF - Osservatorio Astronomico di Brera, Via Bianchi 46, I-23807 Merate, Italy \label{INAFB} 
\and
Max-Planck-Institut fur Physik, D-80805 Munchen, Germany \label{MPIM} 
}

\offprints{H.E.S.S., NuSTAR and Fermi~collaborations,
\protect\\\email{\href{mailto:contact.hess@hess-experiment.eu}{contact.hess@hess-experiment.eu}};
\protect\\\protect\footnotemark[1] Corresponding authors
}

\date{Received ; Accepted}

\abstract {The results of the first ever contemporaneous multi-wavelength observation campaign on the BL Lac object \pks\ involving \swift, \nustar, \fla\ and \hess\ are reported. The use of these instruments allows us to cover a broad energy range, important for disentangling the different radiative mechanisms. The source, observed from June 2013 to October 2013, was found 
in a low flux state with respect to previous observations but exhibited highly significant flux variability in the X-rays. The high-energy end of the synchrotron spectrum can be traced up to 40 keV without significant contamination by high-energy
emission. A one-zone synchrotron self-Compton model was used to reproduce the broadband flux of the source for all the observations presented here but failed for previous observations made in April 2013. A lepto-hadronic solution was then explored to explain these earlier observational results. 
}

\titlerunning{Multi-wavelength observations of \pks}

\maketitle

\section{Introduction}

Blazars are active galactic nuclei (AGN) with an ultra-relativistic jet pointing towards the Earth. The spectral energy distribution (SED) of blazars exhibits two distinct bumps. The low-energy part (from radio to X-ray) is attributed to synchrotron emission while there is still debate on the emission process responsible for the high-energy bump (from X-ray up to TeV). Synchrotron self-Compton (SSC) models reproduce such emission invoking only leptons. The photons are then produced via synchrotron emission and Inverse-Compton scattering. Hadronic blazar models, in which the high-energy component of the blazar SED is ascribed to emission by protons in the jet, or by secondary leptons produced in p-$\gamma$ interactions, have been widely studied \citep[see e.g.][]{1993A&A...269...67M,2000NewA....5..377A,2001APh....15..121M} as an alternative to leptonic models. They have the benefit that they provide a link between photon, cosmic-ray, and neutrino emission from AGNs, and thus open the multi-messenger path to study AGN jets as cosmic-ray accelerators. The interest in hadronic blazar models has recently increased with the first hint (at 3 $\sigma$ level) of an association of an IceCube high-energy neutrino with the flaring $\gamma$-ray blazar TXS 0506+056 \citep{2018Sci...361.1378I}.

To distinguish between the different models, accurate and contemporaneous observations over a wide energy range are of paramount importance. This is possible in particular with the \textit{Nuclear Spectroscopic Telescope Array} (\nustar), launched in 2012, which permits more sensitive studies above 10 keV than previous X-ray missions.
Its sensitivity in hard X-rays up to 79 keV enables an examination of the high-energy end of the synchrotron emission even in high-frequency 
peaked BL Lac (HBL) objects. Such emission is produced by electrons with the highest Lorentz factors, which could be responsible for the \gr\ emission above tens of GeV that can be detected by ground-based facilities such as the High Energy Stereoscopic System (\hess).

One of the best-suited objects for joint observations is \pks\ \citep[$z=0.116$, ][]{1993ApJ...411L..63F}, a well-known southern object, classified as an HBL already with HEAO-1 observations in X-rays \citep{1979ApJ...229L..53S}. The source is a bright and variable \gr\ emitter. Variability with a time scale of about one month
 was reported in the GeV energy range by the \fer-Large Area Telescope (LAT) \citep{2015ApJS..218...23A} as well as day time scale \citep{2009ApJ...696L.150A} and rapid flaring events \citep{2014ATel.6148....1C,2013ATel.4755....1C} . First detected at TeV energies by \citet{1999ApJ...513..161C} in 1996 with the Durham Mark 6 atmospheric Cerenkov telescope, \pks\ has been regularly observed by \hess\ since the beginning of \hess\ operations, allowing detailed studies of the source variability \citep{2017A&A...598A..39H,2019MNRAS.484..749C}. The TeV flux of the object exhibits log-normal flux variability behaviour across the whole energy range \citep{2017A&A...598A..39H,2019MNRAS.484..749C} making its flux level and variability unpredictable with possible huge flaring events in TeV \citep{2007ApJ...664L..71A}. 

An interesting aspect of this object is the fact that several authors \citep{2008ApJ...682..789Z,2008A&A...484L..35F,2016ApJ...831..142M} reported possible contamination of the hard X-ray spectra by the high-energy component (named hard tail hereafter), but unfortunately, no very high-energy (VHE, E$>$100GeV) data were taken at that time to further constrain the VHE \gr\ flux. In the past, only one multi-wavelength campaign with X-ray instruments, \fla, and \hess\ was conducted \citep{2009ApJ...696L.150A}. The gathered data were equally well reproduced either by a leptonic model such as the SSC model \citep{2009ApJ...696L.150A} or a lepto-hadronic model \citep{2012AIPC.1505..635C}. 

\pks\ was then the target of a multi-wavelength campaign from June to October 
2013 by \nustar, \hess, as well as the \textit{Neil Gehrels Swift} Observatory and the \fla. These instruments 
observed \pks\  to provide contemporaneous data for the first time in a very broad energy range, extending from ultra-violet up to TeV \grs\ and  yielding a more complete 
coverage in the X-ray and $\gamma$-ray ranges than the 
previous campaign held in 2008 \citep{2009ApJ...696L.150A}.

This paper presents the gathered multi-wavelength data and the analysis in Section~\ref{datana}. In Section~\ref{discusion}, the variability of the source and the X-ray spectra are discussed. Section~\ref{model} presents the modelling of the data, and Section~\ref{conclusion} summarizes the findings of this campaign.

\section{Data analysis}\label{datana}

\pks\ is an important calibration source in X-rays and was observed during a cross-calibration campaign with other X-ray 
instruments early in the \nustar\ mission \citep{1538-3881-153-1-2}. The multi-wavelength observations of the source in April 2013 including \nustar, \textit{XMM-Newton}, and \fla\ were reported by \citet{2016ApJ...831..142M}, and those are denoted as epoch 0 in this paper. 

\begin{figure*}[htp]
\centering
\includegraphics[width=.90 \textwidth]{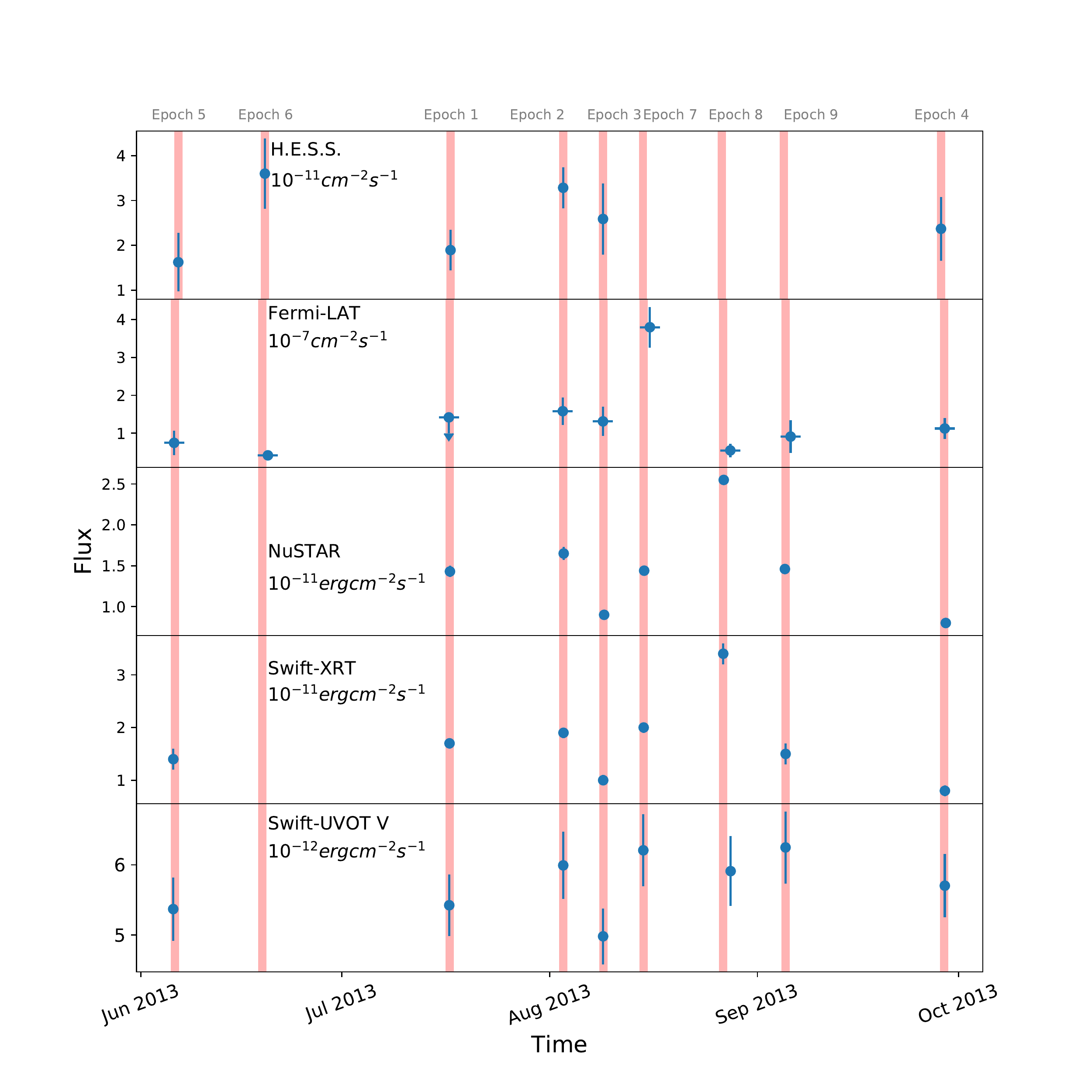}

\caption{Multiwavelength light curve of \pks\ in (from top to bottom) TeV, GeV, X-ray, and UV. The red lines illustrate the epochs mentioned in the text.}
\label{LC}
\end{figure*}

Observations of \pks\ were a part of the ``Principal Investigator'' phase of the \nustar\ mission.  
The aim was to have those observations take place in exact coincidence with observations 
by the $\gamma$-ray observatory \hess\ Due to diverse constraints (technical problems, 
bad weather, etc), \hess, \nustar\ and \swift\ only observed \pks\ 
simultaneously during four epochs, where each epoch corresponds 
to observations conducted on a given night, (2013-07-17, 2013-08-03, 2013-08-08 and 
2013-09-28):  those are labelled as epochs 1, 2, 3 and 4. \hess\ and \swift\ observed 
the blazar for two additional epochs (2013-06-05 and 2013-06-19, labelled 5 and 6). Epoch 6 is presented, for sake of completeness, 
since the \swift\ data were found to be not usable (see Section~\ref{xrtobs}). \nustar\ and \swift\ also 
observed \pks\ during three extra epochs (labelled 7, 8 and 9): those are reported 
here also for the sake of completeness. For each epoch, \fla\ data were analysed and the results are reported in Section~\ref{Fermiobs}. Fig.~\ref{LC} presents the overall light curve derived from all the epochs.

\subsection{\hess\ data analysis and results}

The \hess\ array is located in the Khomas Highland,
in Namibia (23\dgr16'18'' S, 16\dgr30'01'' E), at an altitude of 1800 meters above sea
level. \hess, in its second phase, is an array of five imaging Atmospheric Cherenkov  telescopes. 
Four of the telescopes (CT1-4) have segmented optical reflectors of 12 m diameter consisting of 382 mirrors
 \citep{2003APh....20..111B} and cameras composed of 960 photomultipliers. They form the array of the H.E.S.S. phase I. The second phase started in September 2012 with the addition of a 28 m diameter telescope (CT5) with a camera of 2048 photomultipliers in the centre of the array.

The system operates either in Stereo mode, requiring the
detection of an air shower by at least two telescopes \citep{2004APh....22..285F,2015arXiv150902902H} or 
in Mono mode in which the array triggers on events detected only with CT5.

\pks\ was observed by the full \hess\ phase II array during the present observational
campaign. Table \ref{table:hessobs} gives the date of each observation 
and the results of the analysis described in the following Sections. To ensure good 
data quality, each 28-minute observation had to pass standard 
quality criteria \citep{aha2006}. For two nights (2013-08-03 and 2013-09-28, epochs 2 and 4), 
these criteria have not been met by the four 12 m telescopes. Therefore, only CT5 Mono observations are available for these nights.

Data for each night have been analysed independently using the
{\it Model} analysis \citep{Naurois} adapted for the five-telescope array (hereafter named
Stereo analysis). In this case, {\tt Loose cuts} (with a threshold of 40
photo-electrons) were used to lower the energy threshold. For the Mono
analysis, {\tt standard cuts} (threshold of 60 photo-electrons) were applied to minimize 
systematic uncertainties.

The spectra obtained at each epoch were extracted using a forward-folding method described
in \citet{2001A&A...374..895P}. For each night, a power-law model of the form $\phi_{\rm dec} (E/E_{\rm dec})^{-\Gamma}$, where $E_{\rm dec}$ is the decorrelation energy, was used. Table \ref{table:hessobs} lists the parameters providing the best fits to the data above an energy threshold $E_{\rm th}$. This threshold is defined as the energy where the acceptance is 10\% of the maximal acceptance.

For completeness, the spectra averaged over the epochs 1, 3, 5 and 6 (Stereo mode observations) and over epochs 2 and 4 (Mono mode observations) were computed separately. Above 200~GeV, both measurements are compatible with each other, with a integrated flux of $(4.86 \pm 0.30) \cdot 10^{-6}\,{\rm ph}\,{\rm cm^{-2}}\,{\rm s^{-1}}$ for the Stereo mode observations and  $(2.59 \pm 0.38) \cdot 10^{-6}\,{\rm ph}\,{\rm cm^{-2}}\,{\rm s^{-1}}$ for Mono mode observations. All the \hess\ data have been analyzed together by combining the Stereo and Mono mode observations \citep[see][]{2015arXiv150902902H}, allowing us to compute an averaged spectrum (see Table~\ref{table:hessobs}). The integrated flux above 200~GeV measured for this combined analysis is $(3.12 \pm 0.47) \cdot 10^{-12}\,{\rm ph}\,{\rm cm^{-2}}\,{\rm s^{-1}}\,{\rm TeV^{-1}}$. A cross check with a different analysis chain \citep{2014APh....56...26P} was performed and yields similar results.

\begin{table*}[htp]
\caption{\hess\ observations of \pks. The first five columns give the epoch label,
the observation date,  the
live time, the observation mode and the
energy threshold. The data were fitted with a simple power-law with differential flux 
$\phi_{\rm dec}$  at $E_{\rm dec}$ (the decorrelation energy) and with an
index $\Gamma$. The integrated flux above $E_{\rm th}$ is also given.}

\label{table:hessobs}
\centering
\begin{tabular}{c c c c c c c c c }
\hline\hline 
 epochs& date &live time & Mode & $E_{\rm th}$ & $\phi_{\rm dec}(E_{\rm dec})$  & $\Gamma$ & $E_{\rm dec}$ & Flux\\
 & & [h] & & [TeV] &  [$10^{-12}\,{\rm cm^{-2}}\,{\rm s^{-1}}\,{\rm TeV^{-1}}$]  &  & [TeV] &[$10^{-12}\,{\rm cm^{-2}}\,{\rm s^{-1}}$]  \\
\hline 
1 &2013-07-17 &  1.2 & Stereo & 0.108 & 68.1 $\pm$ 5.5  & 2.89 $\pm$ 0.12 & 0.27 &57.6 $\pm$ 5.4 \\
2 &2013-08-03 &  2.0 & Mono & 0.072 & 324.8 $\pm$ 27.7  & 2.84 $\pm$ 0.14 & 0.18 &173.4 $\pm$  17.2 \\
3 &2013-08-08 &  0.4 & Stereo & 0.120 & 98.9 $\pm$ 11.6  & 2.82 $\pm$ 0.21 & 0.26&59.1 $\pm$ 7.5 \\
4 &2013-09-28 &  1.2 & Mono & 0.072 & 211.5 $\pm$ 28.5  & 2.72 $\pm$ 0.23 & 0.20 &133.4 $\pm$ 20.9\\
5 &2013-06-05 &  0.9 & Stereo & 0.146 & 61.8 $\pm$ 12.3  & 3.17 $\pm$ 0.60 & 0.26 &27.0 $\pm$ 5.8 \\
 6 &2013-06-19 & 0.8 & Stereo & 0.108 & 123.1 $\pm$ 9.1  &  2.79 $\pm$ 0.13 & 0.26 &90.1 $\pm$ 7.8\\
\hline 
Stack &  & 6.5 & Combined & 0.121 & 75.7 $\pm$ 2.7  &  3.00 $\pm$ 0.06 & 0.29 & 62.0 $\pm$2.6  \\
\hline 

\end{tabular}
\end{table*}

\subsection{\fla\ data analysis and results}\label{Fermiobs}

The \fla\ is a \gr\ pair conversion telescope
\citep{2009ApJ...697.1071A}, sensitive to \grs\ above 20~MeV.
The bulk of LAT observations are performed in an all-sky survey mode ensuring a
coverage of the full sky every 3 hours.

Data and software used in this work ({\tt Fermitools}) are publicly available from the Science
Support Center\footnote{\url{https://fermi.gsfc.nasa.gov/ssc/data}}. Events
within 10\dgr\ around the radio coordinates of \pks\ (region of interest, ROI) and passing the
{\tt SOURCE} selection \citep{2012ApJS..203....4A} were considered corresponding to event class 128 and event type 3 and a maximum zenith angle of 90\dgr. Further cuts
on the energy (100~MeV$<$E$<$500~GeV) were made, which remove the events with poor energy resolution. To ensure a significant detection of \pks, time windows of 3 days centred
on the campaign nights (Table \ref{table:hessobs}) were considered to extract 
the spectral parameters. To
analyse LAT data, {\tt P8R3\_SOURCE\_V2} instrumental response functions (irfs)
were used. In the fitting procedure, {\tt FRONT} and {\tt BACK} events \citep{2009ApJ...697.1071A} were treated separately.


The Galactic and extragalactic background models, designed for the {\tt PASS 8}
irfs denoted {\tt gll\_iem\_v07.fits} \citep{2016ApJS..223...26A} and {\tt iso\_P8R3\_SOURCE\_V2\_v1.txt} were used in the sky
model, which also contains all the sources of the fourth general \fer\ catalogue \citep[4FGL,][]{2019arXiv190210045T} within the ROI
plus 2\dgr\ to take into account the large point spread function (PSF) of the instrument especially at low energy.

An unbinned maximum likelihood analysis \citep{1996ApJ...461..396M}, implemented in the {\tt gtlike}
tool\footnote{An unbinned analysis is recommended for small time bins \url{https://fermi.gsfc.nasa.gov/ssc/data/analysis/scitools/binned_likelihood_tutorial.html}.}, was used to find the best-fit spectral parameters of each epoch. Models other than the power-law reported here do not improve the fit quality significantly. Table~\ref{table:fermiobs} shows the results of the analysis. Note that for epoch 1 with a test statistic (TS) below 25 ($\approx 5\sigma$), a flux upper limit was derived assuming a spectral index of $\Gamma=1.75$\footnote{This value has been taken a priori and close to the index found in this work.}.

All the uncertainties presented in this section are statistical only. The most important source of systematic uncertainties in the LAT results is the uncertainty on the effective area, all other systematic effects are listed on the FSSC web site \footnote{\url{https://fermi.gsfc.nasa.gov/ssc/data/analysis/LAT_caveats.html}}.

\begin{table*}[htp]
\caption{\fla\ observations of \pks. The epoch number is given in the first column and the corresponding date in the second.
Other columns present the results of the analysis: TS, differential flux at the 
decorrelation energy, the spectral index $\Gamma$, the decorrelation energy, and integrated flux between 100~MeV and 500~GeV.}
\label{table:fermiobs}
\centering
\begin{tabular}{c c c c c c c }
\hline\hline 
epochs & date & TS & $\phi_{\rm dec}(E_{\rm dec})$  & $\Gamma$ & $E_{\rm dec}$ & Flux\\
     &    &  &  [$10^{-12}\,{\rm cm^{-2}}\,{\rm s^{-1}}\,{\rm MeV^{-1}}$] & & [MeV] & [$10^{-8}\,{\rm ph}\,{\rm cm^{-2}}\,{\rm s^{-1}}$]\\

\hline 

1 & 2013-07-17 &  19.8 & & & & $<$14.2\\
2 & 2013-08-03 & 131.1 & 16.2 $\pm$ 3.4 & 1.99 $\pm$ 0.17 & 909 & 15.8 $\pm$  3.6\\
3 & 2013-08-08&99.8 & 18.5 $\pm$ 5.4 & 2.01 $\pm$ 0.26 & 845 & 13.1 $\pm$ 3.8\\
4 & 2013-09-28&154.6 & 9.3 $\pm$ 1.7  & 1.79 $\pm$ 0.13 & 1280 & 11.3 $\pm$ 2.8\\
5 & 2013-06-05&57.8 & 5.6 $\pm$ 1.5 & 1.93 $\pm$ 0.22 & 1260 & 7.5 $\pm$ 3.2\\
6 &  2013-06-19 & 127.0 &  0.9  $\pm$ 0.3 &  1.38 $\pm$ 0.14  & 4340 &  4.2 $\pm$  1.4 \\
7 & 2013-08-14 &295.1 &  124.0 $\pm$ 14.8  &  2.07 $\pm$ 0.10  & 540  &  39.0 $\pm$ 5.4 \\
8 &2013-08-26  & 163.1 &  1.1  $\pm$ 0.3  & 1.48   $\pm$   0.14 &  3990 &  5.5 $\pm$ 1.8 \\
9 &  2013-09-04 &46.1 &  6.5 $\pm$ 1.8   &   2.02 $\pm$  0.26 &  1160 & 9.1$\pm$ 4.3  \\ \hline
Stack &  &875.0 & 23.4e-11  $\pm$  1.8   &  1.89 $\pm$  0.06  & 1300  &  12.5 $\pm$ 1.6  \\

\hline 
\end{tabular}
\end{table*}

\subsection{\nustar\ data analysis and results}\label{sec:nustar}

The \nustar\ satellite, developed in the NASA Explorer program, features two multilayer-coated 
telescopes, which focus the reflected X-rays onto pixellated CdZnTe focal plane modules and provide an image of 
a point source with the half-power diameter of $\sim 1'$ \citep[see][for more details]{2013ApJ...770..103H}. The advantage of \nustar\ over other X-ray missions is its broad bandpass, 3–-79 keV with spectral resolution of $\sim 1$~keV.  

Table \ref{table:nustarobs} provides the details of individual \nustar\ pointings: this includes the amount of on-source 
time (after screening for the South Atlantic Anomaly passages and Earth occultation) and mean net (background-subtracted) count rates.   

After processing the raw data with the \nustar\ Data Analysis Software (NuSTARDAS) package 
{\tt v1.3.1} (with the script {\tt nupipeline}), the source data were extracted from 
a region of $45"$ radius centred on the centroid of X-ray emission, while the 
background was extracted from a $1.5'$ radius region roughly $5'$ south-west of the source location, located on the same chip.  The choice of these parameters is dictated by 
the size of the point-spread function of the mirror. However, the derived spectra depend very weakly on the sizes of the extraction regions.  The spectra 
were subsequently binned to have at least 30 total counts per re-binned channel.  Spectral channels corresponding nominally to the 3--60 keV 
energy range, in which the source was robustly detected, were considered.  The resulting 
spectral data were fitted with a power-law, modified by 
the Galactic absorption with a column density of $1.7 \times 10^{20}$ atoms cm$^{-2}$ \citep{1990ARA&A..28..215D}, 
using {\tt XSPEC v12.8.2}, with the standard instrumental response matrices and 
effective area derived using the ftool {\tt nuproducts}. The alternate N$_H$ measurement by \citet{2005A&A...440..775K} 
of $1.4 \times 10^{20}$ cm$^{-2}$ was tested, and the best-fit spectral parameters of the source were entirely 
consistent with results obtained by using \citet{1990ARA&A..28..215D} values. Data for both \nustar\ 
detectors were fitted simultaneously, allowing an offset of the normalization factor for the focal plane module B (FPMB) with respect to module FPMA.  Regardless of the adopted models,
the  normalization offset was less than 5\%.  The resulting fit parameters are given in 
Table \ref{table:nustarobs}. More complex models for fitting to the datasets obtained during joint \nustar\ and \swiftxrt\ pointings were considered, 
and those are discussed in Section~\ref{sec:ht}.  

\begin{figure*}[htp]
\centering
\includegraphics[width=.90 \textwidth]{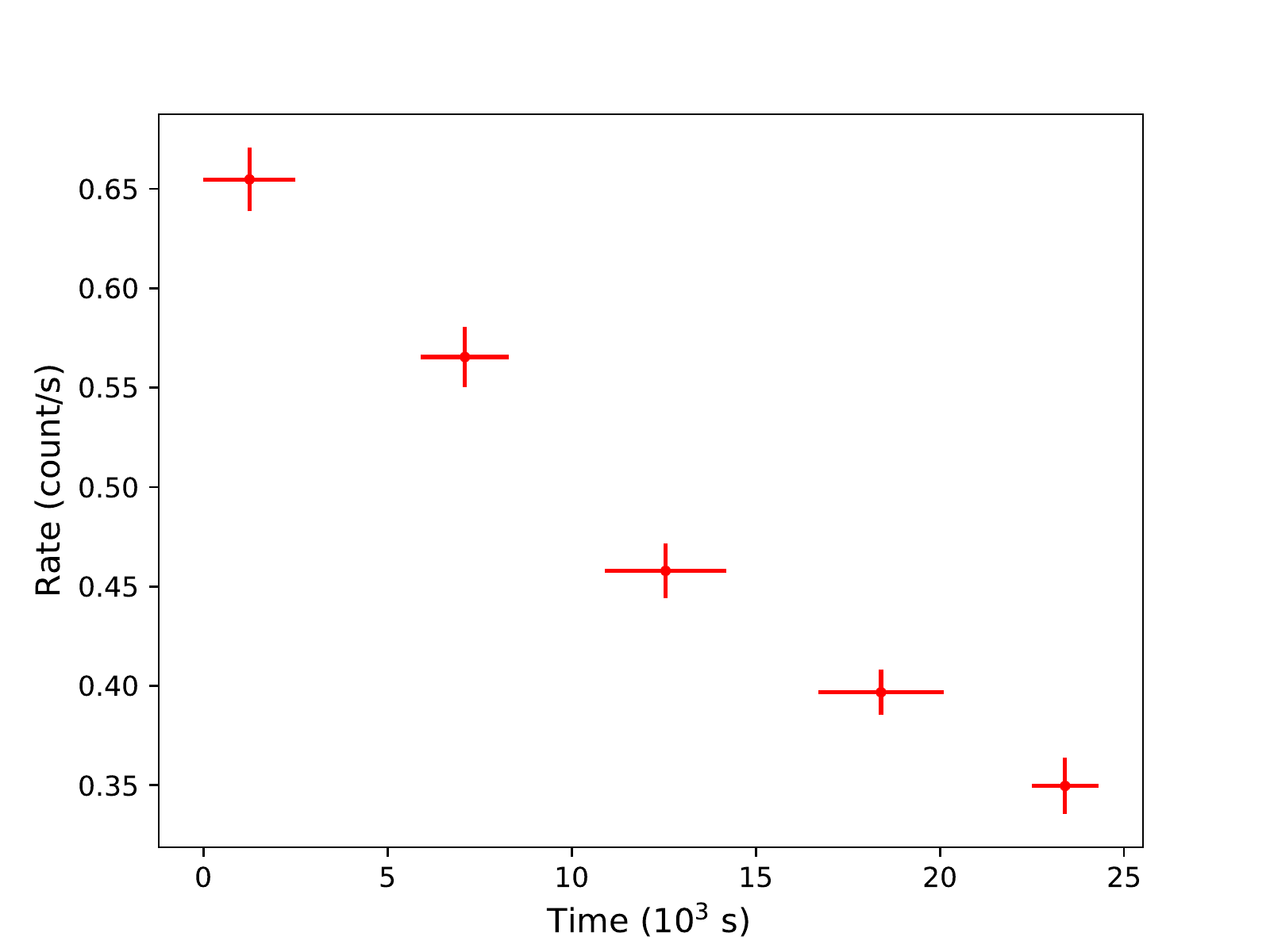}

\caption{Light curve of \pks\ as seen by the FPMA module of \nustar\ during the observation 60002022012 (epoch 8). The energy 
range is 3--60 keV, and the plotted data are not background-subtracted.  However, the background 
rate is always lower than 0.03 counts per second and the background was steady (within 5\%) throughout the observation. Each point corresponds to data taken over roughly one orbit, during the time indicated by the red markers.}
\label{NightAug26}
\end{figure*}

The source exhibited significant variability in one of the pointings, 
on August 26 (epoch 8); the \nustar\ X-ray count rate for the FPMA module 
dropped by almost a factor of 2 in 
25 ks clock time (Fig. \ref{NightAug26}). This was observed independently by both \nustar\ modules.  
The other \nustar\ observations showed only modest variability, with the nominal 
min-to-max amplitude less than 20\% of the mean count rate.  
Such variability is not uncommon in HBL-type BL Lac objects and it has been seen in previous observations of \pks\  
\citep[see, e.g., ][]{2008ApJ...682..789Z}. More recently, rapid X-ray variability was 
seen in \pks\ when it was simultaneously observed by many X-ray instruments \citep{1538-3881-153-1-2}. Other  HBL-type 
blazars exhibit similar variability;  recent examples are  Mkn421 \citep{2016ApJ...819..156B} and  Mkn 501 \citep{2015ApJ...812...65F}.

\subsection{\swiftxrt\ data analysis and results}
\label{xrtobs}

The details of the \swift\ X-ray Telescope \citep[XRT,][]{2005SSRv..120..165B}
observations used here are listed in Table \ref{table:swiftobs}. The observations 
were taken simultaneously (or as close as possible) 
to the \hess\ and \nustar\ observations. During this campaign, \swift\ 
observed the source nine times, but for one of the pointings (corresponding to epoch 6, 
archive sequence 00030795110), applying standard data quality 
cuts resulted in no useful source data (the source was outside of the 
nominal Window Timing -WT- window). Two \swiftxrt\ observations (sequences 0080280006 and -08) 
were close in time and were performed during a single \nustar\ observation. Because these observations
have consistent fluxes and spectra, they were added together as \swiftxrt\ data for epoch~7.

All \swiftxrt\ observations were carried out using the WT 
readout mode. The data were processed with the XRTDAS software package 
(version~3.4.0) developed at Space Science Data Center (SSDC\footnote{\url{https://swift.asdc.asi.it/}}) and distributed by HEASARC within the 
HEASoft package (version~6.22.1). Event files were calibrated and cleaned 
with standard filtering criteria with the {\tt xrtpipeline} task 
using the calibration files available in the \swift\  CALDB (v. 20171113). The average 
spectrum was extracted from the summed cleaned event file. Events for 
the spectral analysis were selected within a circle of 20-pixel 
($\sim$46\arcsec) radius, which encloses about 80\% of the PSF, 
ccentredon the source position. The background was extracted from 
a nearby circular region of 20-pixel radius. The ancillary response 
files (ARFs) were generated with the {\tt xrtmkarf} task applying 
corrections for PSF losses and CCD defects using the cumulative 
exposure map. The latest response matrices (version~15) available 
in the \swift\ CALDB were used. Before the spectral fitting, the 
0.4--10~keV source spectra were binned to ensure a minimum of 30 counts 
per bin. The data extending to the last bin with 30 counts were used, 
which is typically $\sim 5$ keV.

The spectrum of each \swiftxrt\ observation was fitted with a simple power law with a Galactic 
absorption column of $1.7 \times 10^{20}$ atoms cm$^{-2}$, using the {\tt XSPEC v12.8.2} package.  
The resulting mean count rates, power law indices, and corresponding 2--10 keV model 
fluxes are also included in Table \ref{table:swiftobs}.  No variability was found in individual observations in this energy range.

\begin{landscape}
\begin{table*}[htp]

\caption{Summary of the \nustar\ observations of \pks. The first columns are the epoch number, start and stop time of the observation and the corresponding ID. 
The exposure, the count rate of each module and the derived spectral parameters (integrated model flux and photon index) are given in subsequent columns. The last column is 
the $\chi^2$ over the number of bins (Pulse Height Amplitude, PHA). For the power 
law model, the number of degrees of freedom is two less than the number of PHA bins.}
\label{table:nustarobs}
\centering
\begin{tabular}{c c c c c c c c c c}
\hline\hline 
Epoch & Start & Stop & Obs. ID & Exposure & Mod A & Mod B & Flux$_{2-10 {\rm keV}}$ & $\Gamma$  & $\chi^2$/PHA\\
 &  & &  & [ks] & ct rate & ct rate & [$10^{-11}\,{\rm erg}\,{\rm cm^{-2}}\,{\rm s^{-1}}$] &   \\
\hline 
1 & 2013-07-16 22:51:07 & 2013-07-17 07:06:07  & 60002022004 & 13.9 & 0.245 & 0.235 & $1.43 \pm 0.07$ & $2.61\pm 0.05$ &248.3/269\\
2 & 2013-08-02 21:51:07 & 2013-08-03 06:51:07  & 60002022006 & 10.9 & 0.247 & 0.234 & $1.65 \pm 0.08$ & $3.09\pm 0.05$ &188.0/216\\
3 & 2013-08-08 22:01:07 & 2013-08-09 08:21:07  & 60002022008 & 13.4 & 0.149 & 0.133 & $0.90 \pm 0.05$ & $2.85\pm 0.08$ &153.8/159\\
4 & 2013-09-28 22:56:07 & 2013-09-29 06:26:07  & 60002022016 & 11.5 & 0.149 & 0.119 & $0.80 \pm 0.06$ & $2.73\pm 0.07$ &139.1/141\\
\hline\hline
7 & 2013-08-14 21:51:07 & 2013-08-15 07:06:07  & 60002022010 & 10.5 & 0.229 & 0.213 & $1.44 \pm 0.06$ & $2.92 \pm 0.07$ &188.8/195\\
8 & 2013-08-26 19:51:07 & 2013-08-27 03:06:07  & 60002022012 & 11.3 & 0.452 & 0.427 & $2.55 \pm 0.06$ & $2.64 \pm 0.04$ &314.8/333\\
9 & 2013-09-04 21:56:07 & 2013-09-05 07:06:07  & 60002022014 & 12.2 & 0.251 & 0.228 & $1.46 \pm 0.06$ & $2.80 \pm 0.05$ &208.8/238\\
\hline 
\end{tabular}\newline


\caption{Summary of the \swiftxrt\ observations of \pks. The first columns are the epoch number, the start and stop time of the observation and the corresponding ID. The observation length, the count rate and the derived spectral parameters (integrated model flux and photon index) are given in subsequent columns. The last column is the $\chi^2$ over the number of PHA bins (PHA). For the power 
law model, the number of degrees of freedom is two less than the number of PHA bins.}
\label{table:swiftobs}
\centering
\begin{tabular}{c c c c c c c c c }
\hline\hline 
Epochs &Start & Stop & Obs. ID & Exposure & Ct. rate &  Flux$_{2-10 {\rm keV}}$ & $\Gamma$  & $\chi^2$/PHA \\
       &      &         &  &[ks]         &  [cts/s]   & [$10^{-11}\,{\rm erg}\,{\rm cm^{-2}}\,{\rm s^{-1}}]$ &            \\
\hline 
1 & 2013-07-17 00:06:58 & 2013-07-17 02:41:34 & 00080280001 & 1.6 & 1.67 & $1.7 \pm 0.1$ & $2.43 \pm 0.06$ &  79.0/77  \\
2 & 2013-08-03 00:20:59	& 2013-08-03 02:50:45 & 00080280002 & 2.1 & 2.56 & $1.9 \pm 0.1$ & $2.63 \pm 0.05$ & 118.2/124 \\
3 & 2013-08-08 23:06:59 & 2013-08-09 00:21:47 & 00080280003 & 1.7 & 1.36 & $1.0 \pm 0.1$ & $2.71 \pm 0.07$ & 64.8 / 65 \\
4 & 2013-09-28 22:50:59	& 2013-09-29 00:06:47 & 00080280015 & 1.6 & 1.07 & $0.8 \pm 0.1$ & $2.69 \pm 0.08$ & 40.8 / 53 \\
5 & 2013-06-05 19:37:59	& 2013-06-05 20:43:12 & 00030795109 & 0.9 & 1.61 & $1.4 \pm 0.2$ & $2.57 \pm 0.09$ & 45.4 / 45 \\
\hline\hline
7 & 2013-08-14 23:15:45 & 2013-08-15 02:13:48 & 00080280006 and -08 & 1.8 & 2.32 & $2.0 \pm 0.1$ & $2.59 \pm 0.05$ &  89.2 / 108 \\
8 & 2013-08-26 20:17:59	& 2013-08-26 23:06:38 & 00080280009 & 1.0 & 3.1 & $3.4 \pm 0.2$ & $2.38 \pm 0.06$  &  68.1 / 85 \\
9 & 2013-09-05 04:33:59	& 2013-09-05 05:39:41 & 00080280013 & 0.9 & 0.85   & $1.5 \pm 0.2$ & $2.65\pm 0.10$ &  17.2 / 28 \\
\hline 
\end{tabular}
\end{table*}
\end{landscape}

\subsection{Spectral fitting of X-ray data and the search for the hard X-ray ``tail''} \label{xray}

The results of the spectral fits of the \swiftxrt\ and \nustar\ data separately are given in Table \ref{table:nustarobs} and Table \ref{table:swiftobs} respectively.  However, because \pks\ exhibited complex X-ray spectral structure measured in the joint XMM-\textit{Newton} plus \nustar\ observation
in  April 2013 \citep{2016ApJ...831..142M}, here, a joint 
fit to the lower-energy \swiftxrt\ and the higher-energy \nustar\ data was performed to investigate 
the need for such more complex models. Since the source is highly variable, only the strictly simultaneous \swiftxrt\ and \nustar\ data sets were paired.  
To account for possible effects associated with variability or imperfect \swiftxrt\ -- to -- \nustar\ 
cross-calibration, the normalizations of the models for the two 
detectors were allowed to vary, but the difference was in no
  case greater than 20\%, consistent with the findings of \citet{1538-3881-153-1-2}, with the exception of the August 26 observation (epoch 8) where \nustar\ revealed 
significant variability (see note in Section~\ref{sec:nustar}).

To explore the spectral complexity similar to that seen in April 2013, the following models were considered\footnote{Models are corrected for Galactic absorption.} :
(1) PL: a simple power-law model; and  
(2) LP:  a log-parabola model. The resulting joint spectral fits are given in Table \ref{table:nustarswiftobs}.  

In four observations (epochs 1, 3, 4 and 9), the model consisting of a simple PL absorbed by the Galactic column fits the data well: no deviation from a simple power-law model is required. However, for epochs 2, 7 and 8, a  
significant improvement ($\Delta \chi^{2}>20$ for one extra parameter) of the fit quality is found by adopting the LP model. Thus, at these epochs, the spectrum steepens with energy.  In conclusion, there are not only {\sl spectral index} changes from one observation epoch to another, but there is also a significant change of the {\sl spectral curvature} 
from one observation to another. \citet{2018A&A...619A..93B}, using only \nustar\ data, reported results on the same observations and also found a change in the spectral shape for epoch 8 but not for epoch 2 and 7. They also reported a hardening for epochs 1, 3 and 4, but one which is not significant when comparing with a PL fit.

A third model consisting of one log-parabola plus a second hard power-law with spectral index $\Gamma_{\rm HT}$ (LPHT)\footnote{The formula for this LPHT model is $\phi\propto E^{-\Gamma - \beta\cdot log(E)} + E^{-\Gamma_{\rm HT}}$} has also been tested. The model adds a generally harder high-energy ``tail'' (HT) to the softer log-parabola component.  A notable feature is the absence of such a ``hard tail'' in any of the observations (see Section~\ref{sec:ht}). Therefore, an upper limit on the 20--40 keV flux has been computed assuming $\Gamma_{\rm HT} = 2$.

\begin{table*}
\caption{Joint \nustar\ and \swiftxrt\ observations of \pks. The errors quoted on the spectral parameters as well as the quoted 20--40 keV flux limits are 90\% level confidence regions. For the log-parabola model, the number of degrees of freedom is four less than the number of PHA bins, 
since the LP model has one extra parameter, and in addition, the normalization 
of the two instruments is fitted separately. The 2--10 keV flux for joint \textit{Swift} and \nustar\ spectral fits is essentially the same 
as that measured by \nustar\ alone.}
\label{table:nustarswiftobs}
\centering
\begin{tabular}{c c c c c c c c }
\hline\hline 
Epochs  & PL index & $\chi_{\rm PL}^2$/PHA &   LP index &  LP curvature &   $\chi_{\rm LP}^2$/PHA  & Flux$_{\rm HT} (20-40 \mathrm{keV})$ \tablefootmark{b}  \\
              &  $\Gamma$ &     &$\Gamma$\tablefootmark{a}   & $\beta$ & & [$10^{-12}\,{\rm erg}\,{\rm cm^{-2}}\,{\rm s^{-1}}$] \\
\hline 
1  & $2.54 \pm 0.04$ & 341.3 / 346 & $2.57^{+0.13}_{-0.03}$ & $0.13 \pm 0.06$ & 332.1 / 346  & $< 1.2$ \\
2  & $2.80 \pm 0.03$ & 414.0 / 340 & $3.01^{+0.12}_{-0.04}$ & $0.27 \pm 0.07$ & 301.7 / 340  & $< 0.4$ \\
3  & $2.77 \pm 0.05$ & 223.5 / 224 & $2.82 \pm 0.07$     & $0.09 \pm 0.06$ & 218.9 / 224  & $< 0.8$ \\
4  & $2.71 \pm 0.06$ & 179.5 / 194 & $2.71 \pm 0.06$     & $0.00 \pm 0.07$ & 179.5 / 194  & $< 0.8$ \\
\hline\hline
7  & $2.72 \pm 0.04$  & 327.8 / 303 & $2.86^{+0.11}_{-0.05}$ & $0.18^{+0.08}_{-0.04}$ & 281.6 / 303  & $< 0.5$ \\
8  & $2.56 \pm 0.03 $ & 425.1 / 418 & $2.59 \pm 0.03 $  & $0.17 \pm 0.04$ & 378.2 / 418  & $< 0.8 $ \\
9  & $2.78 \pm 0.05$  & 229.5 / 266 & $2.78 \pm 0.05$   & $0.10 \pm 0.15$ & 226.7 / 266  &$< 1.3 $   \\
\hline 
\end{tabular}

\tablefoot{
\tablefoottext{a}{$\Gamma$ is evaluated at 5 keV. }
\tablefoottext{b}{The hard tail index is assumed to have $\Gamma_{\rm HT}$ of 2.}
} 
\end{table*}

\subsection{\swiftuvot\ data analysis and results}

The Ultraviolet/Optical Telescope \citep[UVOT;][]{2005SSRv..120..165B} on board \swift\ also observed \pks\ during \swift\ pointings.
UVOT measured the UV and optical emission in the bands \textit{V} (500--600 nm), \textit{B} (380--500 nm), \textit{U} (300--400 nm), \textit{UVW1} (220--400 nm), \textit{UVM2} (200--280 nm) and \textit{UVW2} (180--260 nm). The values of \citet{2011ApJ...737..103S} were used to correct for the Galactic absorption\footnote{see \url{https://irsa.ipac.caltech.edu/applications/DUST/index.html} with a reddening ratio $A_\mathrm{v}/ E(B-V)=$3.1 and $E(B-V)=$0.022.}.

The photon count-to-flux conversion is based on the UVOT calibration \citep[Section 11
of][]{Poole2008}. A power-law spectral index $\Gamma_{\rm UV}$ has been derived for each epoch and reported in Table \ref{table:uvotobs}. The results presented in this work do not provide evidence for spectral variability in the UV energy range.

\begin{table*}
\caption{\swiftuvot\ observations of \pks. The fluxes are given in units of $10^{-12}\,{\rm erg}\,{\rm cm^{-2}}\,{\rm s^{-1}}$. The last column 
is the power-law spectral index $\Gamma_{\rm UV}$ obtained by fitting the UVOT data.}
\label{table:uvotobs}
\centering
\begin{tabular}{c c c c c c c c }
\hline\hline 
Epochs &\textit{V}&\textit{B} &\textit{U}&\textit{UVW1}&\textit{UVM2}&\textit{UVW2} & $\Gamma_{\rm UV}$ \\
 &\textit{2.30 eV}&\textit{2.86 eV} &\textit{3.54 eV}&\textit{4.72 eV}&\textit{5.57 eV}&\textit{6.12 eV} &  \\
\hline 

0 \tablefootmark{*}  & 71  $\pm$ 2 & 73  $\pm$ 2&78  $\pm$ 3&  75  $\pm$ 3&88  $\pm$ 3& 81  $\pm$ 3 & \\\hline
1  & 54.2  $\pm$  1.5  & 56.0  $\pm$  1.2  & 59.6  $\pm$  1.4  & 59.4  $\pm$  1.2  & 67.1  $\pm$  1.4  & 60.1  $\pm$  1.1  &1.86 $\pm$   0.14\\
2 &59.9  $\pm$  1.6  & 65.4  $\pm$  1.4  & 66.5  $\pm$  1.5  & 69.5  $\pm$  1.4  & 79.5  $\pm$  1.6  & 71.1  $\pm$  1.3  & 1.80 $\pm$   0.14 \\
3 & 49.8  $\pm$  1.3  & 54.4  $\pm$  1.1  & 51.5  $\pm$  1.2  & 57.8  $\pm$  1.1  & 64.9  $\pm$  1.3  & 62.1  $\pm$  1.1  &  1.77   $\pm$ 0.14 \\
4 & 57.0  $\pm$  1.4  & 60.5  $\pm$  1.2  & 61.4  $\pm$  1.4  & 62.9  $\pm$  1.2  & 72.3  $\pm$  1.4  & 63.1  $\pm$  1.1  &  1.86   $\pm$  0.14\\
5 & 53.7  $\pm$  1.6  & 58.5  $\pm$  1.4  & 65.3  $\pm$  1.6  & 64.7  $\pm$  1.4  & 75.8  $\pm$  1.6  & 65.7  $\pm$  1.2  & 1.76 $\pm$  0.14 \\
\hline\hline
7 &62.1  $\pm$  1.8  & 64.3  $\pm$  1.5  & 73.3  $\pm$  1.8  & 74.3  $\pm$  1.5  & 84.5  $\pm$  1.9  & 74.6  $\pm$  1.4  & 1.76 $\pm$  0.14 \\
8 &59.1  $\pm$  1.8  & 60.7  $\pm$  1.5  & 65.6  $\pm$  1.6  & 70.1  $\pm$  1.5  & 79.4  $\pm$  1.7  & 70.1  $\pm$  1.3  &  1.76  $\pm$ 0.14 \\
9 &62.5  $\pm$  1.8  & 68.6  $\pm$  1.6  & 68.0  $\pm$  1.6  & 70.6  $\pm$  1.5  & 81.4  $\pm$  1.7  & 72.6  $\pm$  1.4  & 1.83 $\pm$  0.14\\
\hline 
\end{tabular}
\tablefoot{
\tablefoottext{*}{Values taken from \citet{2016ApJ...831..142M}. }
}
\end{table*}

\section{Discussion}\label{discusion}
\subsection{Flux state and variability in \grs}

During the observation campaign, \pks\ was found in a low flux state, in the \hess\ energy range, $\phi(E>200{\rm GeV})= (11.6 \pm1.3) \times 10^{-12}\,{\rm ph}\,{\rm cm^{-2}}\,{\rm s^{-1}}$, a factor $\approx 5$ lower than during the 2008 campaign 
\citep[$\phi(E>200{\rm GeV}) = (57.6\pm 1.8) \times 10^{-12}\,{\rm ph}\,{\rm cm^{-2}}\,{\rm s^{-1}}$]{2009ApJ...696L.150A}, see Fig. \ref{fig:seds}. 
The average flux above 200~GeV measured by \hess\ during 9 years of observations \citep[$\phi(E>200{\rm GeV}) = (51.0\pm 4.1) \times 10^{-12}\,{\rm ph}\,{\rm cm^{-2}}\,{\rm s^{-1}}$,][]{2017A&A...598A..39H} is also more than 4 times higher than the one reported here (for the entire campaign). Note that even lower flux values have been measured over the last 10 years \citep[see Figure 1 of ][]{2017A&A...598A..39H}.
The source exhibits a harder spectrum ($\Gamma \approx 2.8$) with respect to the \hess\ phase I measurement \citep[$\Gamma \approx 3.4$, ][]{2009ApJ...696L.150A,2017A&A...598A..39H}. 
This is consistent with the results of \citet{2017A&A...600A..89H} and likely to be due to the lower energy threshold achieved with CT5.

The \fla\ flux averaged over the 9 epochs was lower than the one measured in the 3FGL, $(12.6
\pm 0.4) \times 10^{-8}\,{\rm ph}\,{\rm cm^{-2}}\,{\rm s^{-1}}$, and lower than in 2008 by a factor $\approx 2$. Similar results were found by \citet{2017A&A...600A..89H} showing that the source was in a low flux state in 2013. With a flux of $(8
\pm 2) \times 10^{-8}\,{\rm ph}\,{\rm cm^{-2}}\,{\rm s^{-1}}$ in the 100~MeV-300~GeV energy range, epoch 0 is not different from the epochs reported here.

The $2-10$ keV X-ray flux was found to be a factor $\approx 3-4$ lower than in 2008 \citep{2009ApJ...696L.150A}; see  Fig.~\ref{fig:seds}. Only at two epochs (3 and 4), the $2-10$ keV flux measured by \nustar\ was lower than the one measured at epoch 0 ($1.1\times 10^{-11}\,{\rm erg}\,{\rm cm^{-2}}\,{\rm s^{-1}}$) and the fluxes of epochs 1, 2, 7, 8 and 9 were higher.

The only noticeable difference is at lower energies with the observed optical flux measured by \swiftuvot: At epoch 0, the flux was higher than that measured in all the other epochs (see Table~\ref{table:uvotobs}).  

\subsection{Broad-band X-ray spectrum}\label{sec:ht}

In the energy range from 0.3~keV to 10~keV, the spectrum is usually assumed to be the high-energy
end of the synchrotron emission. Indeed, the measured spectral index of \pks\ in the X-ray regime is generally in agreement with the value expected
for a HBL, for which a power-law spectral index, $\Gamma$, is typically 
steeper than 2 (hereafter ``soft component''). Nevertheless a single power law is too simple a representation
of the spectrum when measured with sensitive instruments affording a good signal-to-noise ratio.  
As already pointed out by \citet{0004-637X-625-2-727}, 
the soft X-ray spectra of HBLs are well represented as gradually steepening functions towards higher energies.  
In the data presented here, the spectral index measured by \swiftxrt\ is always harder than the one measured by  \nustar.
A Kolmogorov-Smirnov test was performed on both \swiftxrt\ and \nustar\ spectral index distributions. It rejects the hypothesis that they are sampled from the same distribution with a P-value of 3\%. This suggests that such steepening takes place for \pks.

At the end of the X-ray spectrum (roughly above a few keV), \citet{1982ApJ...253...38U}
observed \pks\ above an energy of a few keV with the HEAO A1 instrument,
 and \citet{2008ApJ...682..789Z} reported a hard excess in two XMM-\textit{Newton} observations (confirmed by
\citet{2008A&A...484L..35F} using the same observations). The XMM-\textit{Newton} observations fit with a broken power-law showed a spectral
hardening of $\Delta \Gamma = 0.1-0.3$ with a break energy of 3--5~keV. Both works interpreted this as a possible contamination of the synchrotron spectra by inverse-Compton emission.

More recently, and with the increased energy range provided by \nustar, 
\citet{2016ApJ...831..142M} also measured a hard tail in the X-ray spectrum of
\pks\ (April 2013 observations, epoch 0). Using a broken power-law model, they found a flattening spectrum with a spectral break of $\Delta
\Gamma > 1 $ around 10~keV. During that observation, the source
was found in a very low flux state (with the 2--10 keV flux of $1.1 \times 10^{-11}$
erg\,cm$^{-2}$\,s$^{-1}$), even lower than the flux reported by
\citet{2008ApJ...682..789Z} and \citet{2008A&A...484L..35F}. Fitting jointly the strictly simultaneous 
XMM-\textit{Newton} data with the \nustar\ data, a more complete picture emerged, with 
a log-parabola describing the soft ($E < 5$ keV) spectrum, and a ``hard tail,'' 
which can be described as an additional power-law.

Concerning the observations presented in this work, adding an extra hard tail (LPHT model) 
does not significantly improve the $\chi^2$. However, it is important to note that the flux of the object 
during the April 2013 pointing was relatively low, and the observations were fairly long (about 4 times 
longer than any single pointing during the campaign reported here).  
As noted by \citet{2016ApJ...831..142M}, the hard tail becomes more easily detectable only 
during low-flux states of the softer, low-energy spectral component. 

To detect a possible hard tail in the data set of the present campaign, 
a spectral fit of all data sets simultaneously was performed. Due to the spectral variability of the 
soft, low-energy component (Table \ref{table:nustarobs}), stacking (or just summing) all 
simultaneously is inappropriate. Instead, a simultaneous fit of seven individual datasets from Epochs 1, 2, 3, 4, 7, 8, and 9, was considered, allowing the spectral parameters of the soft component (described as a log-parabola) to vary independently. Each epoch was described by a LPHT model (see Section~\ref{xray}), and with common normalization of the ``hard tail'' for all data sets\footnote{In an SSC or lepto-hadronic scenario, one would expect the hard X-ray tail to be the low energy counterpart of the \fer\ spectra. The approach made here with the assumption of a constant normalization for the tail is more conservative than using the \gr\ spectral results.}. Formally, the fit returns zero flux for the hard tail component. The 99\% confidence upper 
limit of $1.8 \times 10^{-4}\,{\rm ph}\,{\rm keV^{-1}}\,{\rm cm^{-2}}$ on the normalization of this component (at $\chi^2 + 2.7$) corresponds to 
a 20--40 keV flux limit of $2.5 \times 10^{-13}$ erg~cm$^{-2}$~s$^{-1}$.  
The normalization of this hard tail in epoch 0 data was $8 \times 10^{-4}\,{\rm ph}\,{\rm keV^{-1}}\,{\rm cm^{-2}}$ (corresponding to 
a 20--40 keV flux of $12.0 \times 10^{-13}$ erg cm$^{-2}$ s$^{-1}$), 
or more than 4 times higher than the upper limit measured during the other epochs. In conclusion, the hard tail is 
also variable on the time scale of months, but no conclusions on the shorter time scales 
from the presented \nustar\ data can be drawn.

Note that the source does exhibit a similar flux level in X-rays with respect to the April 2013 data set while in optical, the flux is significantly lower. In an SSC framework, this photon field might be scattered by low energy electrons to produce hard X-ray photons, accounting for the hard tail visible in epoch 0. Nevertheless, when the \fer\ measurement is extrapolated down towards the \nustar\ energy range, it always overshoots the X-ray measurement. This can be due to a lack of statistics in the LAT range preventing the detection of spectral curvature as the one reported in the 3FGL catalog, since only 3 days of data were used in each epoch. The extrapolation of the 3FGL spectrum of \pks\ does not violate the upper limits derived here on the hard tail component but cannot reproduce epoch 0.

\section{SED Modeling}\label{model}
\subsection{Leptonic modelling : one zone synchrotron self-Compton}

Modelling of blazar SEDs was performed with a one-zone SSC model by \citet{THEO::SSC_BAND}. The emission zone is considered to be sphere of radius $R$ 
filled with a magnetic field $B$ and moving at relativistic speed with a Lorentz factor $\Gamma$. In this zone, the emitting particle distribution follows a broken power-law:

\begin{equation}
n_e (\gamma) = 
  \left\{
      \begin{array}{lr}
        N \gamma^{-p_1} & \text{ if $\gamma_{min} < \gamma < \gamma_b$} \\
        N \gamma^{-p_2} \, \gamma_b^{p_2 - p_1} & \text{ if $\gamma_b < \gamma < \gamma_{max}$} 
      \end{array}
    \right.
\end{equation}
where $N$ is density of electrons at $\gamma=1$. $p_1$ and $p_2$ are the indices of the electron distribution and  $\gamma_b$ the break energy.

The modeling was performed for on the epochs presented in this work (1-5) with UV, X-ray, GeV and TeV data. Radio data from \citet{2010ApJ...716...30A} and \citet{2013AJ....145...73L} were taken from the NED\footnote{\url{http://ned.ipac.caltech.edu/}}. The radio emission could originate from another location in the jet, or from the emission zone, and is therefore considered as upper limits in the model. Historical data taken between $10^{-2}$~eV and $1$~eV (Infra-red range) are found to be quite stable in time with variation less than a factor 2. Such data have been collected using Vizier\footnote{\url{http://cds.u-strasbg.fr/vizier-org/licences_vizier.html}} and shown in the SEDs.

\begin{figure*}[t!]
\centering
        \includegraphics[width=0.49\textwidth]{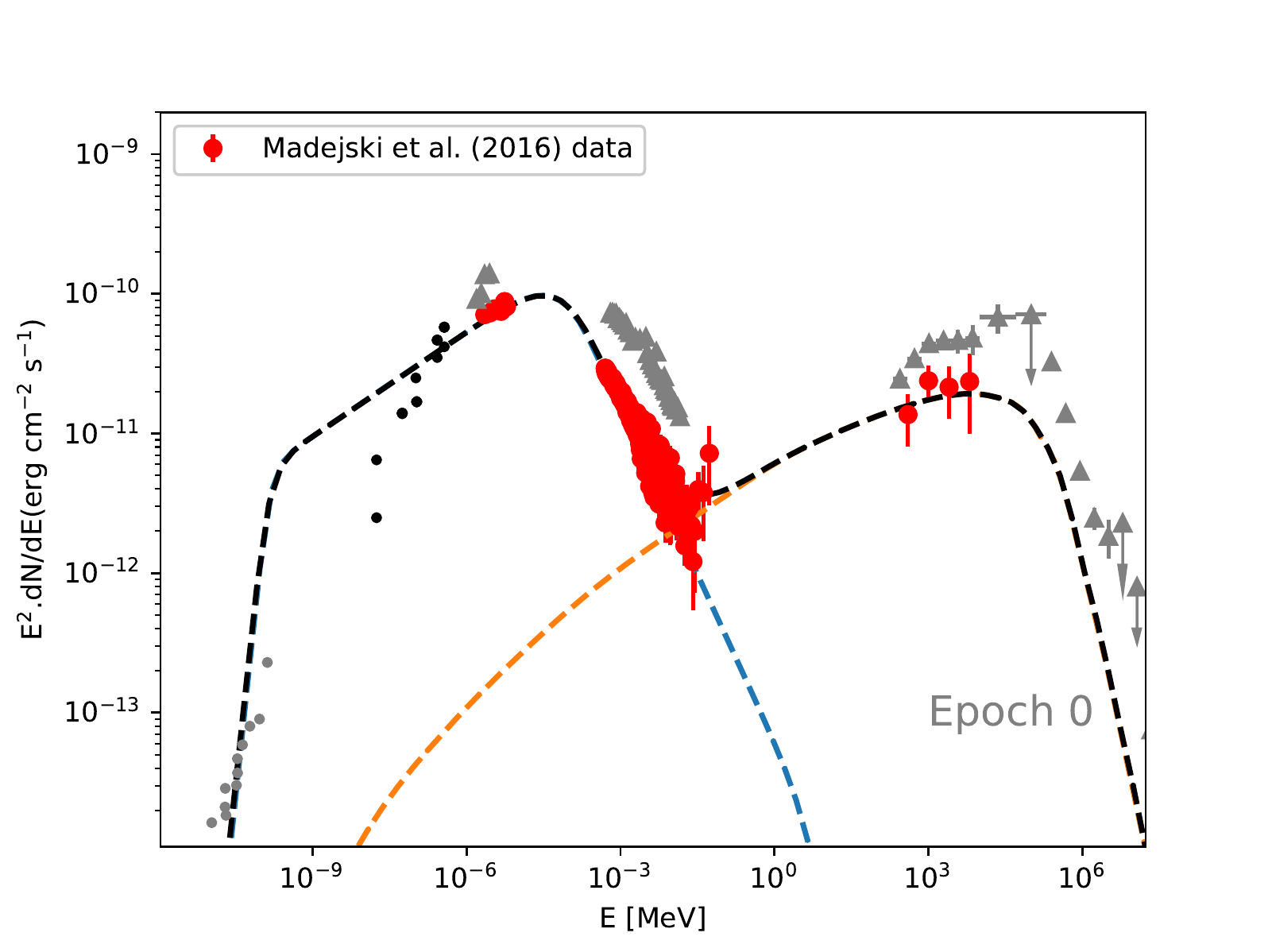}
        \includegraphics[width=0.49\textwidth]{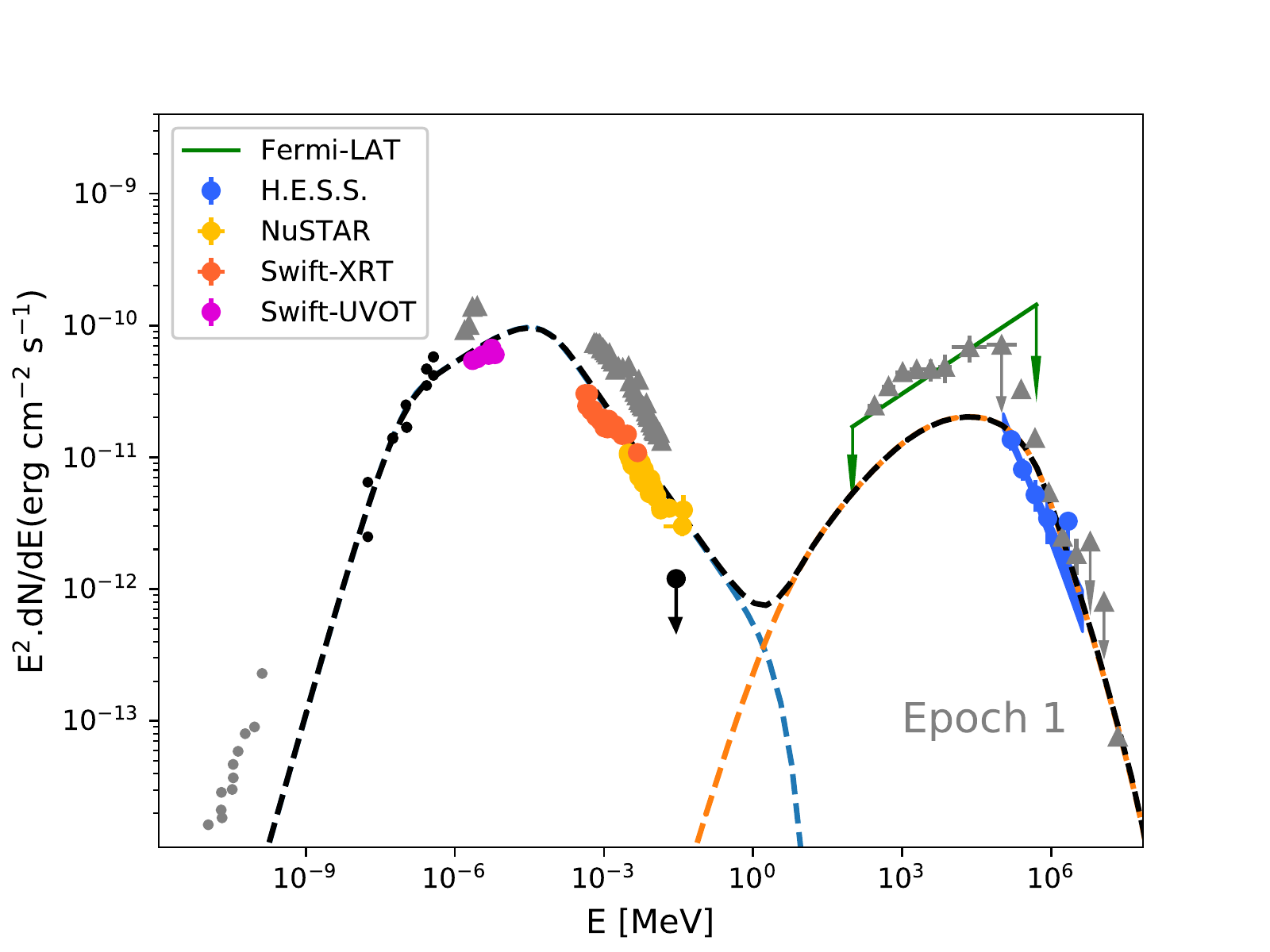}
        \includegraphics[width=0.49\textwidth]{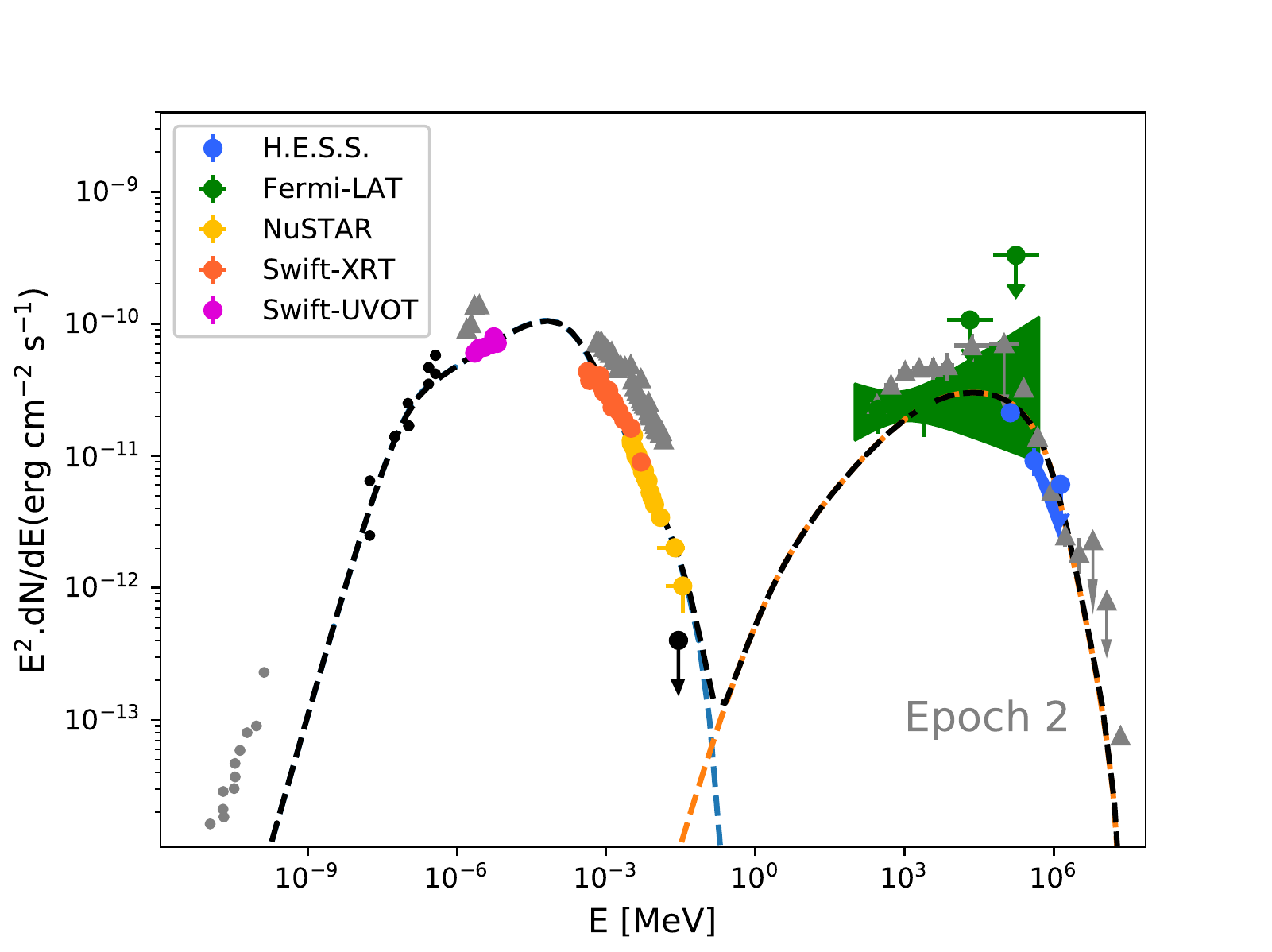}
        \includegraphics[width=0.49\textwidth]{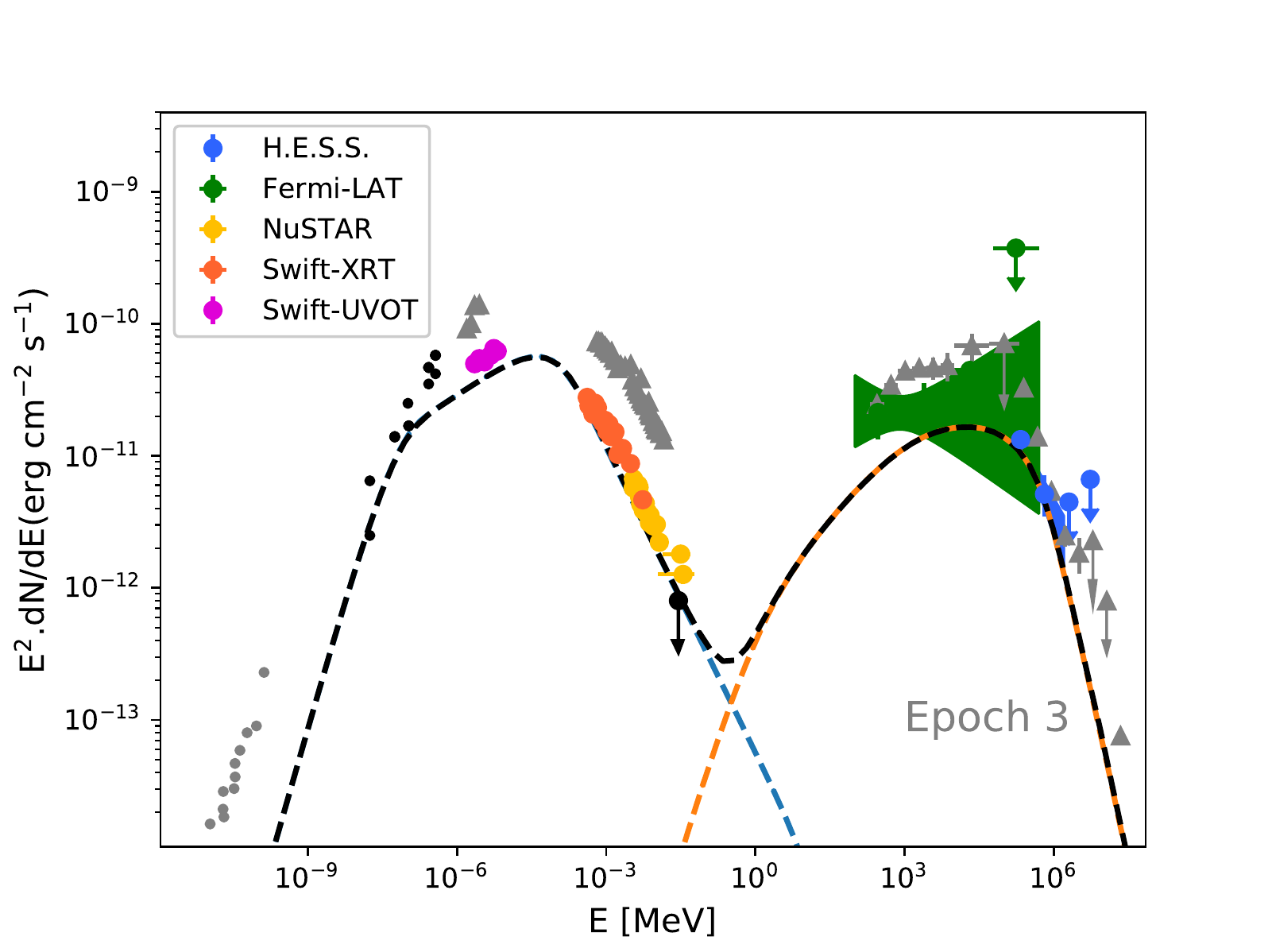}
        \includegraphics[width=0.49\textwidth]{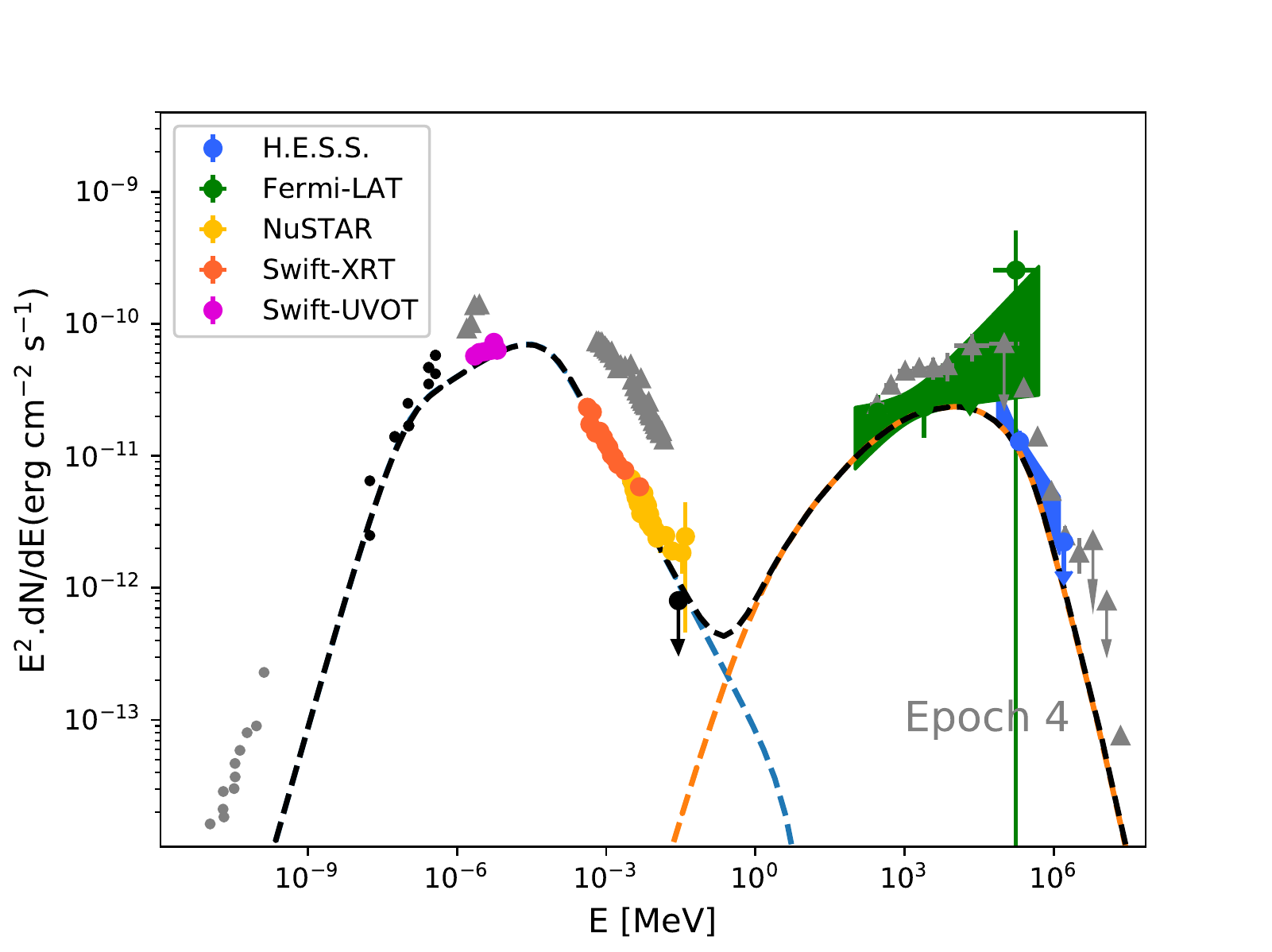}
        \includegraphics[width=0.49\textwidth]{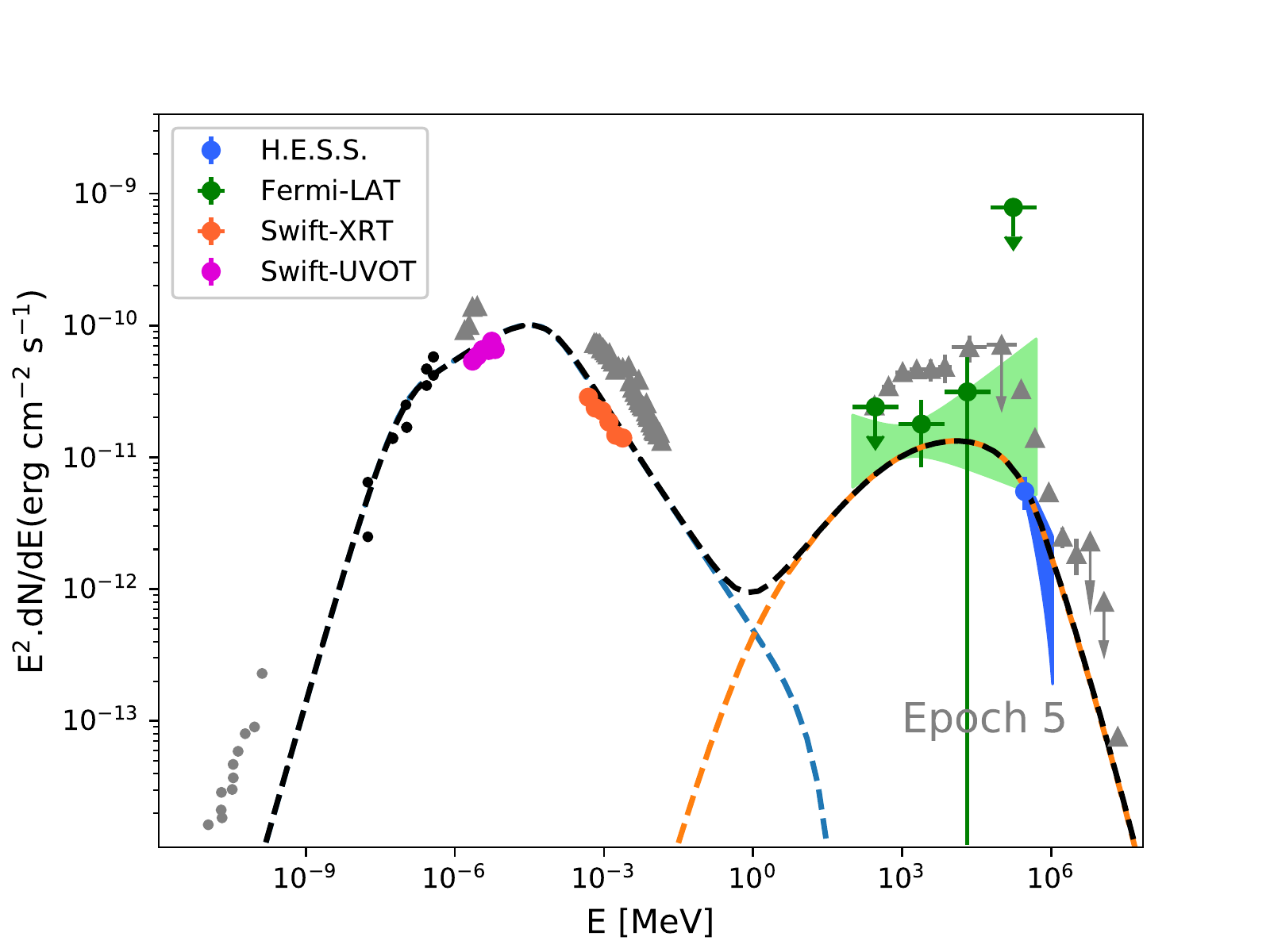}

    \caption{Spectral energy distribution of \pks\ for each epoch considered in this work. For epoch 0, the red points are directly extracted from \citet{2016ApJ...831..142M}. In the other plots, the purple points are UVOT data, orange are XRT data, yellow are the \nustar\ data. In \grs, the green points and contours are the \fla\ results and \hess\ results are in blue. The black upper limits refer to the hard tail component (see text) and is used to constrain the inverse-Compton part of the SSC model (black line). The grey points are the data from the 2008 observation campaign \citep{2009ApJ...696L.150A} shown for comparison. Black points are the radio data from \citet{2010ApJ...716...30A,2013AJ....145...73L}. The dashed blue line is the synchrotron emission, the orange line is the IC emission. Both are from the SSC calculation, and the black dashed line is the sum of both.}
     \label{fig:seds}
\end{figure*}

For each epoch, a mathematical minimization \citep{NelderMead} was performed to find the model parameters $R$, $B$,  $N$, $\log (\gamma_{min})$, $\log (\gamma_b)$ and $\log (\gamma_{max})$ that best fit the data. The values of $p_1$ and $p_2$ were constrained by the UV and X-ray data, respectively, and not let free in the fitting procedure. Given the little spectral variability found in UV and GeV, $p_1$ was set to 2.5 and $p_2=2\cdot\Gamma_{\rm Xray}-1$ \citep{1986rpa..book.....R}. The minimization was performed using a Markov Chain Monte Carlo (MCMC) implemented in the {\tt emcee} python package \citep{2013PASP..125..306F}. For epochs 1-4, the upper limit on the hard tail flux (Table \ref{table:nustarswiftobs}) is taken into account by forcing the inverse-Compton (IC) component of the model to be below this limit. The resulting parameters are given in Table \ref{tab:parameters} with their corresponding realizations in Fig.~\ref{fig:seds}.

The model parameters are consistent with previous studies by \cite{Kataoka:2000jw}, \cite{Foschini:2007ku}, \cite{Katarzynski:2008es} and \cite{Aharonian:2009es}. As in these previous studies, as well as for other BL Lac objects (e.g. Mrk~421
\citep{2011ApJ...736..131A}, Mrk~501 \citep{2011ApJ...727..129A}, SHBL~J001355.9--185406 \citep{2013A&A...554A..72H}, etc...), the obtained model is far from equipartition. Even with a very low flux state in the present modelling, particles carry at least 10 times more energy density than the magnetic field.

The data from epochs 1-5 are well reproduced by the simple SSC calculation presented here. In contrast to \citet{2017ApJ...850..209G} 
for this object or \citet{2017ApJ...842..129C} for Mrk~421, there is no need to invoke a second component to reproduce the SED without over-predicting the radio flux. 
The main difference is that the hard tail above $\approx$ 10 keV seen in the previous observations is not observed in the present data set.

The SSC model was applied to the data of epoch 0 with results also presented in Table \ref{tab:parameters}. The contemporaneous data are well reproduced. The main difference in the modelling parameters between epoch 0 and the campaign presented in this work lies in the values of $\gamma_{min}$. For epoch 0, having $\log (\gamma_{min})=0$ allows a greater inverse-Compton contribution in the X-ray band, making the X-ray tail detectable by \nustar. This is also in agreement with the observed decrease in the optical flux in epochs 1-5. Indeed a higher value of $\gamma_{min}$ decreases the number of electrons emitting in this energy range. Note also that the archival radio data are in disagreement with the modelling of epoch 0, which predict a too high flux in that energy range. The values obtained for different parameters are not equally well constrained. The shape of the electron distribution ($\gamma_{min}$, $\gamma_{break}$ and $\gamma_{max}$) is quite robust with small errors. Other parameters like the B-field or the size of the emitting region remain poorly known and are indeed different from the model presented in \cite{2016ApJ...831..142M}.

\begin{table}[htp]
\begin{center}
\caption{Model parameters for each epoch. Errors were estimated from the MCMC distributions. The first column recalls the epoch, then minimal, break and maximal energies are given, then the indices $p_1$ and $p_2$. The last parameters are the B-field, size of the region $R$ and the total number of electrons $N_{tot}$. The equipartition factor (ratio of the energy carried by electron over energy in the magnetic field $U_e/U_b$) is given in the last column.}
\label{tab:parameters}
\begin{tabular}{ccccccccccc}
\hline \hline
 Epochs &  $\log(\gamma_{min})$ &  $\log(\gamma_b) $&  $\log(\gamma_{max})$ &   $p_1$ &  $p_2$ & $\delta$ &$B$  &   $R$ &  $N_{tot}$ &  $U_e/U_b$\\
 {} &  {} &   {} &  {} &   {} &  {} & &  [$10^{-2}$ G]  &  [$10^{16}$ cm] &   $[10^{+50}]$ &   \\
\hline
  0 &  0.21$^{+0.01 }_{-0.01  }$ & 4.69$^{+0.01 }_{-0.01}$ & 7.09$^{+ 0.11}_{-0.20}$  & 2.5 & 4.60 &  33.0$^{+1.8 }_{-1.7}$ & 4.2$^{+ 0.2}_{-0.3 }$ &  5.9$^{+0.6}_{-0.5}$&   4317.8$^{+322.9}_{-617.9 }$  &  722.0 \\\hline
  1 &  3.55$^{+0.06}_{-0.11}$ & 4.96$^{+0.06}_{-0.08}$ &  7.31$^{+0.43}_{-0.54}$ & 2.5  & 4.10  & 27.1 $^{+ 1.7 }_{-1.5 }$ & 1.2 $^{+0.4  }_{- 0.3 }$ &  24.5 $^{+16.0 }_{-7.7 }$&  5.8$^{+2.6}_{-  2.2 }$ &  11.8 \\
  2 &  3.39$^{+ 0.06  }_{- 0.07  }$ & 5.02$^{+ 0.04 }_{- 0.07 }$ &  6.27$^{+ 0.21 }_{- 0.19}$ & 2.5 & 4.60 & 32.4$^{+ 2.0 }_{-1.5 }$ & 2.0$^{+ 0.3}_{- 0.3 }$ &  10.6$^{+ 2.3 }_{- 5.1 }$ &  2.7 $^{+ 937.2}_{- 0.8 }$ &  18.7 \\
  3 &  3.39$^{+ 0.10 }_{- 0.16 }$& 4.95$^{+ 0.11 }_{- 0.09 }$ &  7.55$^{+ 0.17 }_{- 0.57 }$ & 2.5 & 4.54 & 29.2$^{+ 3.2 }_{-4.1 }$ & 1.7$^{+ 1.2 }_{- 0.7 }$ & 10.8$^{+ 5.1 }_{- 6.5 }$ &  2.9 $^{+ 1.5 }_{- 2.9 }$&   23.4 \\
  4 &  3.32$^{+ 0.11}_{- 0.10 }$ & 4.73$^{+ 0.11 }_{- 0.11}$ &  7.14$^{+ 0.47 }_{- 0.53 }$ & 2.5 & 4.42 & 30.6$^{+ 4.0 }_{-2.3}$ &  3.1$^{+ 1.4}_{- 1.2 }$ &  6.2$^{+ 5.7 }_{- 2.9 }$ & 1.6$^{+ 1.5 }_{- 0.7 }$ &  19.1 \\
  5 &  3.29$^{+ 0.10 }_{- 0.14 }$ & 4.74$^{+ 0.08 }_{- 0.15 }$ &  7.42$^{+ 0.43 }_{- 1.04 }$ & 2.5 & 4.14 & 32.8$^{+ 2.2 }_{-3.4}$ & 2.8$^{+ 2.9 }_{- 0.8 }$ &  7.4$^{+ 0.4 }_{- 1.0 }$&  1.6$^{+ 1.0 }_{- 0.9 }$ & 5.6 \\
\hline
\end{tabular}
\end{center}
\end{table}

\subsection{Emergence of a hadronic component in hard X-rays ?}

Following the detection of a \gr\ flare from TXS~0506+056 in coincidence with a high-energy neutrino \citep{2018Sci...361.1378I}, several authors have independently shown that, while pure hadronic models cannot reproduce the multi-messenger dataset, a scenario in which the photon emission is dominated by an SSC component with a sub-dominant hadronic component is viable \citep[see, e.g.,][]{2018ApJ...863L..10A,2018arXiv180704335C,2018arXiv180704275G,2018ApJ...864...84K}. The hadronic component emerges in the hard-X-rays as synchrotron radiation by secondary leptons produced via the Bethe-Heitler pair-production channel in this scenario. With this result in mind, it was investigated whether the hardening seen in the \nustar\ data of \pks\ could be due to sub-dominant hadronic emission. Starting from the simple SSC model for epoch 0 (see Table \ref{tab:parameters}), a population of relativistic protons was added. It was assumed that $p_p = p_{e,1}$ (i.e., protons and electrons share the same acceleration mechanism, resulting in the same injection spectral index) and that the maximum proton Lorentz factor $\gamma_{p,max}$ is determined by equating acceleration and cooling time-scales. The proton distribution was normalized such that the hadronic component emerges in hard X-rays. For additional details on the hadronic code used see \citet{2015MNRAS.448..910C}. Another change in the SSC part of the model was the increase of the value of $\log(\gamma_{min})$ to 3.3 in order to not overshoot the radio emission.

\begin{figure*}[t!]
\centering
        \includegraphics[width=0.9\textwidth]{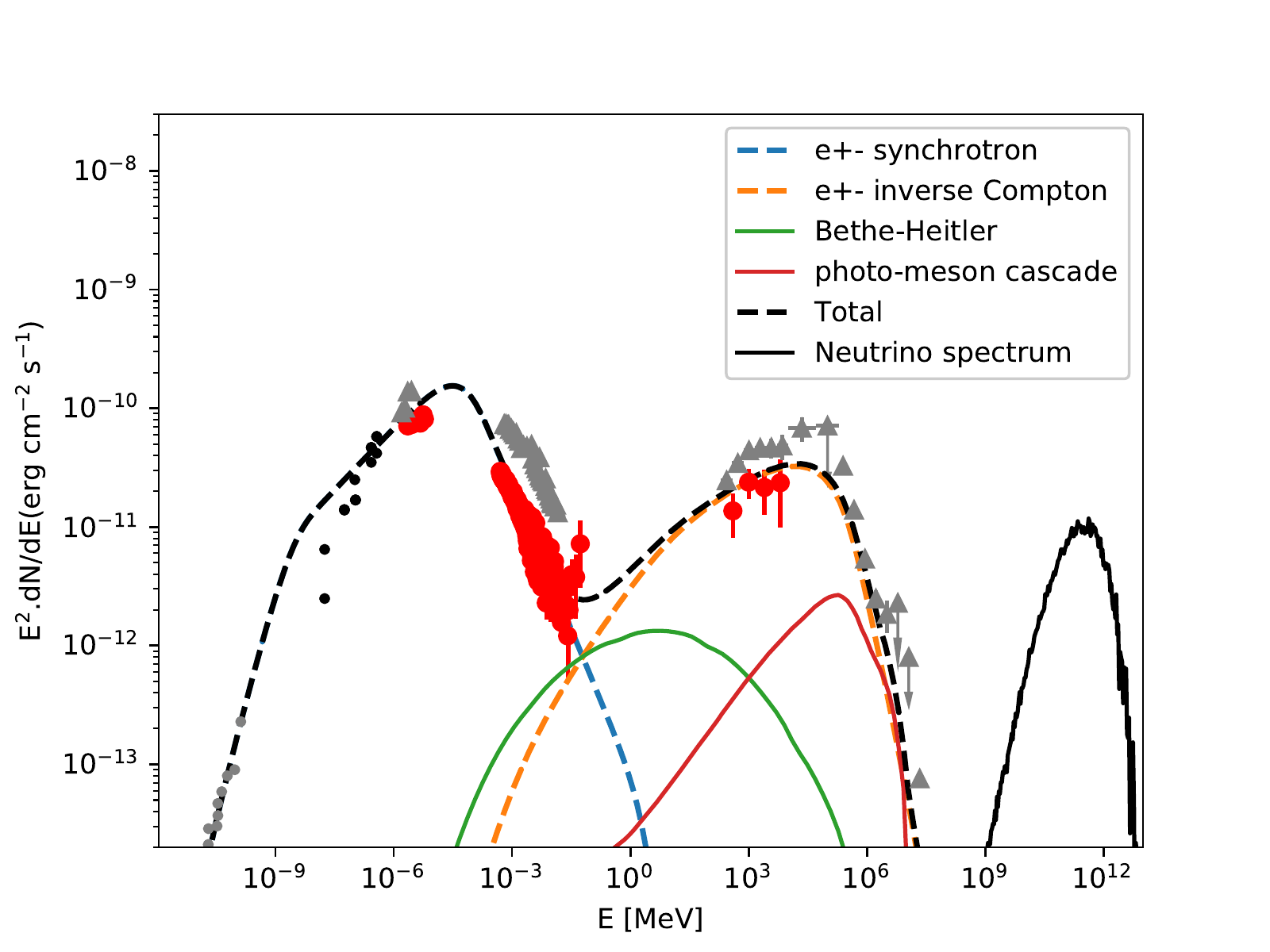}

    \caption{Same as Fig. \ref{fig:seds} but for epoch 0 only. The blue and orange dashed lines are the synchrotron and inverse-Compton emission as in Fig. \ref{fig:seds}. The green line is the emission form Bethe-Heitler pair-production and red from the photo-meson cascade. The sum of all these components is given by the black dashed line. The black continuous line is the predicted neutrino spectrum.}
     \label{fig:sedshadro}
\end{figure*}

The key parameter is the power in protons $L_p$ required to provide the observed photon flux, because a very well-known drawback of hadronic blazar models is that they often require proton powers well above the Eddington luminosity $L_{Edd}$ of the super-massive black hole which powers the AGN. For the case of \pks, if $p_p = 2.3$, $\gamma_{p,min} = 1$ and $\log{\gamma_{p,max}}=8.0$, $L_p = 3.5 \times 10^{49}$ erg s$^{-1}$ is needed, which is around $100 L_{Edd}$ for a black hole mass of $10^9_{M_\odot}$, making this scenario unrealistic. This result is very sensitive to the exact shape of the proton distribution, especially at low Lorentz factors (that cannot be constrained by data).  $L_p$ is lower if the proton distribution is harder, or if $\gamma_{p,min} > 1$. As an example, if $p_p = 2.0$ and $\gamma_{p,min} = 1000$, $L_p = 6.3 \times 10^{47}$\ergs, of the same order of magnitude as $L_{Edd}$.  For this scenario, the hadronic photon emission is shown in Fig. \ref{fig:sedshadro}, and emerges in X-rays as the emission by Bethe-Heitler pairs, and at VHE as photo-meson cascade. The model predicts an expected neutrino rate in IceCube of $\nu_{rate} = 0.03$ yr$^{-1}$, which is compatible with the non-detection of \pks\ by IceCube (computed using the IC effective area\footnote{\url{https://icecube.wisc.edu/science/data/PS-IC86-2011}} for a declination of $-30^\circ$).


\section{Conclusions}\label{conclusion}

\pks\ was, for the first time, observed contemporaneously by \swift, \nustar, \fla\ and \hess The source was found in a low flux state in all wavelengths during epochs 1-9. The source flux is lower than during the campaign carried out in 2008.

For each epoch, no hard tail was detected in the X-ray spectra, contrary to what was seen at epoch 0. The computation of an upper limit on the 20--40 keV flux of such a hard tail for each observation and for the full data set shows that this component is variable on the time scale of a few months. For epochs 1-5, the SED is well reproduced by a one-zone SSC model. Such a model fails to reproduce the epoch 0 data due to the required value of the $\gamma_{min}$ parameter. A low value of $\gamma_{min}$ is mandatory to reproduce the hardening in X-rays but in return produces a too high-flux in the radio band with respect to the archival measurements.

The emergence of the variable X-ray hard tail cannot be explained by a one-zone SSC model. Several authors proposed a multi-zone model to tackle this issue, and especially \citet{2017ApJ...850..209G} used a spine/layer jet structure. In such a structured jet, synchrotron photons of the slow layer are Comptonized by the electrons of the fast spine to produce the hard X-ray tail.  The results presented here would imply that the layer, producing the hard tail is variable on monthly time scales. Such a result is in agreement with their model parameters. Nevertheless, the variability time scale derived from their model parameters cannot reproduce variability of the source on a time scale of days, as the model was not designed to reproduce such variability.

Here the possibility of having a lepto-hadronic radiation component was explored. The same parameters as for the SSC model but with $\log(\gamma_{min}) = 3.3$ were used to reproduce a large part of the SED. The hard tail was successfully reproduced by the hadronic emission. 
Nevertheless for such a model to not be in disagreement with the Eddington luminosity of the super-massive black hole, the proton distribution has to be harder ($p_e=2.0$) than the electron distribution ($p_e=2.3$) together with $\gamma_{p,min}>1000$ and/or have a low-energy cut-off $\gamma_{p,min} > 1$. In the framework of the lepto-hadronic model, the hard-X-ray emission associated with Bethe-Heitler pair production is independent and not directly associated with the electron-synchrotron and the SSC components. The detection of the hard-X-ray tail only during one of the \nustar\ observations can thus be explained by a sudden increase in the hadronic injection.

The origin of the hard tail is still uncertain but this feature could help to disentangle  different classes of emission models for \pks\ and blazars in general.

\begin{acknowledgements}
The support of the Namibian authorities and of the University of Namibia in facilitating 
the construction and operation of H.E.S.S. is gratefully acknowledged, as is the support 
by the German Ministry for Education and Research (BMBF), the Max Planck Society, the 
German Research Foundation (DFG), the Helmholtz Association, the Alexander von Humboldt Foundation, 
the French Ministry of Higher Education, Research and Innovation, the Centre National de la 
Recherche Scientifique (CNRS/IN2P3 and CNRS/INSU), the Commissariat à l’énergie atomique 
et aux énergies alternatives (CEA), the U.K. Science and Technology Facilities Council (STFC), 
the Knut and Alice Wallenberg Foundation, the National Science Centre, Poland grant no. 2016/22/M/ST9/00382, 
the South African Department of Science and Technology and National Research Foundation, the 
University of Namibia, the National Commission on Research, Science \& Technology of Namibia (NCRST), 
the Austrian Federal Ministry of Education, Science and Research and the Austrian Science Fund (FWF), 
the Australian Research Council (ARC), the Japan Society for the Promotion of Science and by the 
University of Amsterdam. We appreciate the excellent work of the technical support staff in Berlin, 
Zeuthen, Heidelberg, Palaiseau, Paris, Saclay, Tübingen and in Namibia in the construction and 
operation of the equipment. This work benefited from services provided by the H.E.S.S. 
Virtual Organisation, supported by the national resource providers of the EGI Federation.

The Fermi LAT Collaboration acknowledges generous ongoing support from a number
of agencies and institutes that have supported both the development and the operation of
the LAT as well as scientific data analysis. These include the National Aeronautics and
Space Administration and the Department of Energy in the United States, the
Commissariat à l'Energie Atomique and the Centre National de la Recherche Scientifique
/ Institut National de Physique Nucléaire et de Physique des Particules in France, the
Agenzia Spaziale Italiana and the Istituto Nazionale di Fisica Nucleare in Italy, the
Ministry of Education, Culture, Sports, Science and Technology (MEXT), High Energy
Accelerator Research Organization (KEK) and Japan Aerospace Exploration Agency
(JAXA) in Japan, and the K. A. Wallenberg Foundation, the Swedish Research Council
and the Swedish National Space Board in Sweden.

Additional support for science analysis during the operations phase from the following
agencies is also gratefully acknowledged: the Istituto Nazionale di Astrofisica in Italy and
and the Centre National d'Etudes Spatiales in France. This work performed in part under DOE Contract DE-AC02-76SF00515.

This  work  was  supported  under  NASA  Contract No. NNG08FD60C  and  made  use  of  data  from  the
NuSTAR mission, a project led by the California Institute of Technology, managed by the Jet Propulsion Laboratory, and
funded by the National Aeronautics and Space Administration. We thank the NuSTAR Operations, Software, and Calibration
teams for support with the execution and analysis of these observations. This research has made use of the NuSTAR Data
Analysis Software (NuSTARDAS) jointly developed by the ASI Science Data Center (ASDC, Italy) and the California Institute of Technology (USA).

 This research has made use of the NASA/IPAC Extragalactic Database (NED) which is operated by the Jet Propulsion Laboratory, California Institute of Technology, under contract with the National Aeronautics and Space Administration.

  This research made use of Enrico, a community-developed Python package to simplify Fermi-LAT analysis \citep{2013arXiv1307.4534S}.

This research has made use of the VizieR catalogue access tool, CDS,
 Strasbourg, France. The original description of the VizieR service was
 published in A\&AS 143, 23
  
This work has been done thanks to the facilities offered by the Université Savoie Mont Blanc MUST computing center.

M.~Cerruti has received financial support through the Postdoctoral Junior Leader Fellowship Programme from la Caixa Banking Foundation, grant  n. LCF/BQ/LI18/11630012

M.B. gratefully acknowledges financial support from NASA Headquarters under the NASA Earth and Space Science Fellowship Program (grant NNX14AQ07H), and from the Black Hole Initiative at Harvard University, which is funded through a grant from the John Templeton Foundation.
\end{acknowledgements}


\bibliography{PKS2155_mwlNustar_HESS}
\bibliographystyle{aa}

\end{document}